\documentclass[12pt,showpacs,preprintnumbers,amsmath,aps,amssymb]{revtex4}
\usepackage{graphicx}
\usepackage{dcolumn}
\usepackage{bm}
\usepackage{color}
\begin{document}

\baselineskip 18pt

\title{Quantum fractionary cosmology: K-essence theory}

\author{J. Socorro$^{1}$ }
\email{socorro@fisica.ugto.mx}
\author{J. Juan Rosales $^2$}
\email{rosales@ugto.mx} \affiliation{$^{1}$ \quad Department of
Physics,  Division of Science and Engineering, University of
Guanajuato, Campus Le\'{o}n,
 Le\'{o}n 37150,  Mexico\\
$^{2}$ \quad Department of Electrical Engineering, Engineering
Division Campus Irapuato-Salamanca, University of Guanajuato,
Salamanca 36885, Mexico}

\begin{abstract}
Using a particular form of the quantum K-essence scalar field, we
show that in the quantum formalism, a fractional differential
equation in the scalar field variable, for some epochs in the
Friedmann--Lema$\hat{\i}$tre--Robertson--Walker (FLRW) model
(radiation and inflation-like epochs, for example), appears
naturally. In the classical analysis, the kinetic energy of scalar
fields can falsify the standard matter in the sense that we obtain
the time behavior for the scale factor in all scenarios of our
Universe by using the Hamiltonian formalism, where the results are
analogous to those obtained by an algebraic procedure in the
Einstein field equations with standard matter.
 In the case of the
quantum Wheeler--DeWitt (WDW) equation for the scalar field $\phi$,
a fractional differential equation of order
$\beta=\frac{2\alpha}{2\alpha-1}$ is obtained. This fractional
equation belongs to different intervals, depending on the value of
the barotropic parameter; that is to say, when $ \omega_X \in
[0,1]$, the order belongs to the interval $1\leq \beta \leq 2$, and
when $ \omega_X \in [-1,0)$, the order belongs to the interval $0<
\beta \leq 1$. The corresponding quantum solutions are also given.


\end{abstract}

  \maketitle                            

\section{Introduction}
In general, fractional calculus (FC) is the natural generalization
of ordinary calculus~\cite{Podlubny}. That is, FC considers
integrals and derivatives of non-integer order. Despite the fact
that there is no fully accepted physical and geometric
interpretation of fractional derivatives, during~the last three
decades, FC has been the subject of intense theoretical and applied
studies, because~different types of fractional derivatives have
emerged in the scientific literature, each with its advantages and
disadvantages~\cite{Ortigueira1}. There are many works that
investigate fractional calculus and its applications~\cite{r-s},
being a powerful mathematical tool for describing complex processes,
such as the tautochrone problem~\cite{Abel}, models based on memory
mechanism~\cite{Caputo}, anomalous diffusion~\cite{Wyss}, linear
capacitor theory~\cite{Westerlund}, non-local description of quantum
mechanics~\cite{Hermann},
 processing of medical images~\cite{Jorge,Leo}, and~so on. FC has been quite successful in many areas of science
 and engineering {\cite{Uchaikin,Tarasov,Magin}.}

With the exception of some fractional models that arise naturally,
for example, in Ref.~\cite{Westerlund}, the~vast majority of models
starts from an ordinary differential equation corresponding to a
certain physical model. Then, it is fractionated; that is,
the~derivatives of the system are taken as fractional by applying
any of the definitions: Riemann--Liouville, Caputo,
Caputo--Fabrizio, and Atangana--Baleanu.

Recently, the~FC has been applied to the general theory of
relativity, as in Ref.~\cite{Roberts}, and in particular, to~the
Friedmann--Lema$\hat{\i}$tre--Robertson--Walker model,
with~interesting results in
cosmology~\cite{co1,co2,co3,co4,co5,co6,co7,co8,co9,co10,co11,co12,paulo3,paulo4,paulo5},
and quantum mechanics to quantum cosmology~\cite{paulo1,paulo2}.
These have been featured recently in~the chapter ``Fractional
Quantum Cosmology in Challenging Routes'' in a Quantum Cosmology
book~\cite{paulo6}, and in the publication revising fractional
cosmology~\cite{genly}. In~a previous paper~\cite{Socorro}, we
mentioned that in the quantum formalism, applied to different epochs
for the K-essence theory, we would get a fractional Wheeler--DeWitt
equation in the scalar field component. Now, we report our results
in this direction. Additionally, one point worth mentioning is that
by employing the classical information on the barotropic parameter
of the scalar field, we present a relation for using a real
parameter, which can reproduce the different epochs of our Universe
(inflation-like phenomenon, radiation era, stiff  and dust matter)
for particular values. With~this parameter, we found that the scale
factor is analogous to those found in a previous work for the
standard matter in the same scenarios~\cite{berbena,cervantes}. In
this sense, we can introduce the idea that the kinetic energy of the
scalar field should falsify the standard matter by employing the
K-essence~formalism.

Usually, K-essence models are restricted to the Lagrangian density
of the form~\cite{1,roland,chiba,bose,arroja,tejeiro}
\begin{equation}
 S=\int d^4x \, \sqrt{-g}\,\left[ f(\phi) \, {\cal G}(X)\right],
\end{equation}
where the canonical kinetic energy is given by ${\cal
G}(X)=X=-\frac{1}{2}\nabla_\mu \phi \nabla^\mu \phi$, $f(\phi)$ is
an arbitrary function of the scalar field $\phi$, and~g is the
determinant of the metric. K-essence was originally proposed as a
model for inflation, and then as~a model for dark energy, along with
explorations of unifying dark energy and dark
matter~\cite{roland,bilic,bento}. Another motivation to consider
this type of Lagrangian originates from string
theory~\cite{string,garriga1}. For~more details on K-essence applied
to dark energy, one can see Ref.~\cite{copeland} and
references~therein.

On another front, the~quantized version  of this theory has not been
constructed, perhaps due to the difficulties in building up the ADM
formalism for it. Thus, we transform this theory to a conventional
one, where the dimensionless scalar field is obtained from an
energy-momentum tensor as an exotic matter component; and in this
sense, we can use this structure for the quantization program, where
the ADM formalism is well-known for different classes of
matter~\cite{ryan1}.

This work is arranged as follows: in Section~\ref{sec1}, we give
some definitions of the fractional calculus that we employ in this
work and the main ideas over the K-essence formalism, applied to
obtain the classical solution to the scalar field, including the
fractional parameter defined in a general way. In
Section~\ref{sec2}, we construct the Lagrangian and Hamiltonian
densities for the plane FLRW cosmological model, considering a
barotropic perfect fluid for the scalar field in the variable X, and
present the general case. Next, the~radiation particular scenario,
where fractional momenta appear in the scalar field,  will be used
in  Section~\ref{sec3}, where we present the quantum regime for
several cases of interest. Finally, Section~\ref{sec4} is devoted
to~discussions.

\subsection{Basic Definitions from Fractional~Calculus}\label{sec1}
 Fractional derivatives are defined by means of analytical continuation of the Cauchy's formula for the multiple integral of an
 integer order as a single integral with a power-law kernel into the field of real order $\gamma >0$, \cite{Podlubny,Uchaikin}.
  The fractional integral of order $\gamma$ is written~as:
\begin{equation}
I^\gamma f(t) = \frac{1}{\Gamma(\gamma)} \int_0^t \frac{f(\tau)}{( t
- \tau)^{1 - \gamma}} d\tau, \label{FI}
\end{equation}
recovering an ordinary integral when $\gamma \to 1$. The~Caputo
fractional derivative of order $\gamma \geq 0$ of a
 function $f(t)$ is defined as the fractional order integral (\ref{FI}) of the integer order derivative (in the following, in~the conventional notation, the sub-index 0 corresponds to the
definition domain (0,x) by example, where other notations appear as
(a,b), a+ and b-. In~other words, they are the derivation limits).
\begin{equation}
\frac{d^\gamma}{dt^\gamma}f(t)={^C}_{0}D^{\gamma}_{t} f(t) = I^{n
-\gamma}\, _{0}D_t^n f(t) = \frac{1}{\Gamma(n - \gamma)}\int_0^t
\frac{f^{(n)}(\tau)}{( t - \tau)^{\gamma - n +1}} d\tau, \label{C1}
\end{equation}
with $n-1 <\gamma \leq n \in \mathbb{N} = {1,2, ...}$, and~$\gamma
\in \mathbb{R}$ is the order of the fractional derivative and
$f^{(n)}$ are the ordinary integer derivatives, and~$\Gamma(x) =
\int_0^\infty e^{-t} t^{x - 1} dt$ is the gamma function. The~Caputo
derivative satisfies the following relations:
\begin{eqnarray}
{^C}_{0}D_t^\gamma [ f(t) + g(t)] &=& {^C}_{0}D_t^\gamma f(t) + {^C}_{0}D_t^\gamma  g(t), \label{C2}\\
{^C}_{0}D_t^\gamma c &=& 0,\qquad {where}\,\, c\,\, { is\,\, a\,\,
constant}. \label{C3}
\end{eqnarray}

The Laplace transform of the function $f(t)$ is defined as
\begin{equation}
\mathbb{L}[ f(t)] = \int_0^\infty f(t) e^{-st} dt = F(s).
\end{equation}
Then, the~Laplace transform of the Caputo fractional derivative
(\ref{C1}) has the form~\cite{Podlubny},
\begin{equation}
\mathbb{L}[{^C}_{0}D_t^\gamma f(t) ] = s^\gamma F(s) -
\sum_{k=0}^{n-1} s^{\gamma - k - 1} f^{(k)}(0), \label{C4}
\end{equation}
where $f^{(k)}$ is the ordinary~derivative.

Another definition which will be used is the Mittag--Leffler
generalized function \cite{libro,ML} (and references therein),
defined as the series expansion; in a general way, under~a Maclaurin
series, with~$z \in \mathbb{C}$ 
\begin{equation} E_{\chi, \sigma}(z) =
\sum_{n=0}^\infty \frac{z^n}{\Gamma(\chi n + \sigma)},\qquad (\chi
>0,\,\, \sigma > 0), \label{C5}
\end{equation}
and for $\sigma=1$, we have one parametric Mittag--Leffler function
\begin{equation}
E_\chi(z)=E_{\chi, 1}(z) = \sum_{n=0}^\infty \frac{z^n}{\Gamma(\chi
n + 1)}, \qquad \chi > 0. \label{C5-1}
\end{equation}

Some special cases are~\cite{libro,ML}:
\begin{eqnarray}
&& E_1(\pm z)=e^{\pm z},\qquad E_2(z)=Cosh\left(\sqrt{z} \right),
\qquad E_2(-z^2)=Cos\left(z \right), \nonumber\\
&& E_{2,2}(z^2)=\frac{Sinh\left(z\right)}{z},\qquad
E_{2,2}(-z^2)=\frac{Sin\left(z\right)}{z}. \label{cases}
\end{eqnarray}

Laplace transform of the Mittag--Leffler function is given by
Equation ~\cite{Podlubny}
\begin{equation}
\int_0^\infty e^{-st} t^{\chi m + \sigma - 1}E_{\chi,\sigma}^{(m)}
(\pm at^\chi) dt = \frac{ m!\, s^{\chi - \sigma}}{(s^\chi \mp a
)^{m+1}}. \label{C7}
\end{equation}

Consequently, the~inverse Laplace transform is
\begin{equation}
\mathbb{L}^{-1}\Big[  \frac{ m!\, s^{\chi - \sigma}}{(s^\chi \mp a
)^{m+1}}  \Big] =
 t^{\chi m + \sigma - 1}E_{\chi,\sigma}^{(m)} (\pm at^\chi).
\end{equation}

This expression is very useful for obtaining analytical~solutions.

\subsection{ K-Essence~Theory   }
One of the  simplest  K-essence Lagrangian densities is
\begin{equation}
 {\cal L}_{geo}=\left( R+  f(\phi) {\cal G}(X)\right),
\label{lagrangian}
\end{equation}
where $R$ is the scalar of curvature, $f(\phi)$ and ${\cal G}(X)$
have been defined before. Then, the~field equations are given by
\begin{eqnarray}
 G_{\mu \nu}+ f(\phi) \left[{\cal G}_X \phi_{,\mu}\phi_{,\nu}
 +  {\cal G} g_{\mu \nu}  \right]
&=& T_{\mu \nu}, \label{Efe} \\
  f(\phi)\left[{\cal G}_X\phi^{,\nu}_{\, ;\nu} + {\cal
G}_{XX}X_{;\nu}\phi^{,\nu} \right] + \frac{df}{d\phi}\left[{\cal G}
- 2X{\cal G}_X\right]&=&0, \label{Sfe}
\end{eqnarray}
where we have assumed the units with $8\pi G=1$ and, as~usual,
the~semicolon means a covariant derivative, and~the subscript $X$
denotes differentiation with respect to $X$. \linebreak (Equations
(\ref{Efe}) and (\ref{Sfe}) are deduced in Appendix~\ref{AppA}).

The same set of Equations~(\ref{Efe}) and (\ref{Sfe}) is obtained if
we consider the scalar field $X(\phi)$ as part of the matter
content, to say $ {\cal L}_{X,\phi}= f(\phi) {\cal G}(X)$, with~the
corresponding energy-momentum tensor
\begin{equation}
 {\cal T}_{\mu \nu}= f(\phi)\left[{\cal G}_X \phi_{,\mu}\phi_{,\nu}
 +  {\cal G}(X)g_{\mu \nu}  \right]. \label{ener-mom}
\end{equation}

Additionally, considering the energy-momentum tensor of a barotropic
perfect fluid,
\begin{equation}
 T_{\mu \nu}=(\rho +P)u_\mu u_\nu + P g_{\mu \nu},
\label{tensor-matter}
\end{equation}
with $ u_\mu$ being the four-velocity satisfying the relation $
u_\mu u^\mu=-1$, $\rho$ the energy density, and $P$ the pressure of
the fluid. For~simplicity, we consider a co-moving perfect fluid.
The~pressure and energy density, corresponding to the energy
momentum tensor of the field $X$, are
\begin{equation}
P(X)= f(\phi) {\cal G}, \qquad \rho(X)= f(\phi)\left[2X{\cal
G}_X-{\cal G} \right]; \label{pX}
\end{equation}
thus, the~barotropic parameter $\omega_X= \frac{P(X)}{\rho(X)} $ for
the equivalent fluid  is
\begin{equation}
 \omega_X=\frac{{\cal G}}{2X{\cal G}_X-{\cal G}}.
\end{equation}

Notice that the case of a constant barotropic index $\omega_X$,
(with the exception when $\omega_X=0$) can be obtained by the
 ${\cal G}$ function
\begin{equation}
{\cal G}=X^{\frac{1+\omega_X}{2\omega_X}}.
\end{equation}

Choosing the barotropic parameter as
\begin{equation}
\omega_X=\frac{2\kappa-1}{2\kappa+1}, \qquad \to \qquad {\cal
G}=X^\alpha, \label{ansats}
\end{equation}
 where the $ \alpha$ parameter
\begin{equation}
 \alpha =\frac{2\kappa}{2\kappa-1},
\label{fc-parameter}\end{equation} is relevant in our approach.
Thus, we can write the barotropic parameter in terms of $
\omega_X=\frac{1}{2\alpha -1}$, when
$\kappa=\frac{\alpha}{2(\alpha-1)}$. With~this, we can write
 the  states
in the evolution of our Universe as:
\begin{equation}
\left\{ \begin{tabular}{lcl} stiff matter :& $\kappa \to \infty,$& $
\omega_X=1$, $\to$
$ {\cal G}(X)=X$. \\
Radiation: & $\kappa=1$,&$ \omega_X=\frac{1}{3}$, $\to$  $ {\cal
G}(X)=X^2$.
\\
Radiation like: & $\kappa=\frac{5}{4}$,&$ \omega_X=\frac{2}{3}$,
$\to$ ${\cal G}(X)=X^{5/2}$.\\ such as dust like: & $\kappa \to
\frac{1}{2}$,& $ \omega_X \to 0$, $\to$ $ {\cal
G}(X)=X^m$, \quad$ {m}\to \infty$.\\
 inflation : &$\kappa=0$, & $ \omega_X=-1$, $\to$ $ {\cal
G}(X)=1,\,\, f(\phi)=\Lambda=constant.$\\
inflation such as  &$\kappa=\frac{1}{4}$, & $
\omega_X=-\frac{1}{3}$,
$\to$ $ {\cal G}(X)=\frac{1}{X}$\\
&$\kappa=\frac{1}{10}$, & $ \omega_X=-\frac{2}{3}$, $\to$ $ {\cal
G}(X)=\frac{1}{\sqrt[4]{X}}.$
\end{tabular}
\right. \nonumber
\end{equation}
The classical and quantum solutions for the stiff matter case $
\omega_X=1$ with the function ${\cal G}(X)=X$  were treated in the
Ref.~\cite{1,Socorro}, considering anisotropic cosmologies, which
are the standard quintessence, such as  $f(\phi)= { constant}$. For
the inflation phenomenon,  we chose the particular value for the
cosmological constant function $f(\phi)=\Lambda$. The original
Einstein field Equations~(\ref{Efe}) were reduced to the traditional
problem with the cosmological constant with exponential time
behavior for the scale factor~\cite{berbena}.

It is clear that the stiff-matter case falls into the traditional
treatment of quintessence cosmology, and in the other cases,
a~Hamiltonian density with a fractional momentum in the scalar
field; then, the~ quantum Wheeler--DeWitt equation appears as a
fractional differential equation. In~the Ref.~\cite{jorge1},
the~authors present the classical analysis of the radiation era by
using dynamic systems and obtaining rebound~solutions.

In the following, by~choosing the generic formula of the barotropic
parameter $\omega_X=\frac{1}{2\alpha-1}$,  we obtain the classical
and quantum~solutions.

\subsection{Classical Cosmological FLRW Model, $f(\phi) = Constant$
} The space-time background to be considered is the spatially flat
FLRW with element
\begin{equation}
ds^2=-N(t)^2 dt^2 + A^2(t) \left[dr^2 +r^2(d\theta^2+sin^2\theta
d\phi^2) \right], \label{frw}
\end{equation}
where $N(t)$ represents the lapse function, $ A(t)=e^{\Omega(t)}$ is
the scale factor in the Misner parametrization, and  $\Omega$ is a
scalar function, whose interval is $ (-\infty,\infty)$. If~we
consider the cosmological FLRW model, then the Equation~(\ref{Sfe})
is written as (we use $\prime=\frac{d}{d\tau}=\frac{d}{Ndt}$, so $
g_{\tau \tau}=-1$),
\begin{equation}
 \left[{\cal G}_X  + 2 X {\cal G}_{XX} \right]X^\prime +
6\frac{A^\prime}{A} X {\cal G}_X =0, \label{frw-k-model}
\end{equation}
with the exact  solution
\begin{equation}
 X{\cal G}_X^2=\eta A^{-6},\label{frw-k-solution}
\end{equation}
where $A$ is the scale factor of the cosmological FLRW model,
and~$\eta$ is an integration constant, which is linked to the
parameters of matter in the Universe epoch in study. This solution
has been known for some time and was found by different
authors~\cite{sol,scherrer1,arroja1}. (Equations (\ref{frw-k-model})
and (\ref{frw-k-solution}) have been deduced in
Appendix~\ref{AppB}.)

In the following, we present the generic case, $
\omega_X=\frac{1}{2\alpha-1}$, given by ${\cal G}=X^\alpha$ and
substituting into (\ref{frw-k-solution}), we have $\alpha^2
X^{2\alpha-1}=\eta A^{-6}$ with
$X=\frac{1}{2}\left(\frac{d\phi}{d\tau}\right)^2$, obtaining for the
scalar field $\phi$ the equation
\begin{equation} \frac{d\phi}{d\tau}=
\sqrt{2}
\left[\left(\frac{\eta}{\alpha^2}\right)^{\frac{1}{2(2\alpha-1)}}
A^{-\frac{3}{2\alpha-1}}\right]=\sqrt{2}
\left[\left(\frac{\eta}{\alpha^2}\right)^{\frac{1}{2(2\alpha-1)}}
e^{-\frac{3}{2\alpha-1}\Omega}\right],\label{phi-tau}
\end{equation}
 whose solution is
\begin{equation}
 \Delta \phi=\sqrt{2}
{\left(\frac{\eta}{\alpha^2}\right)^{\frac{1}{2(2\alpha-1)}}} \int
A^{-\frac{3}{2\alpha-1}} d\tau =\sqrt{2}
{\left(\frac{\eta}{\alpha^2}\right)^{\frac{1}{2(2\alpha-1)}}}  \int
e^{-\frac{3}{2\alpha-1}\Omega} d\tau, \label{ss1}
\end{equation}
which is dependent on the scale factor; it was obtained by using the
relationship between the momenta and the Hamiltonian density
constraint from the Lagrangian density, in~the usual~way.

\section{Hamiltonian Cosmological~Models}\label{sec2}
Introducing the line element (\ref{frw}) in  Misner's
parametrization, the~Ricci scalar becomes $R=-6\frac{\ddot
\Omega}{N^2}-12\left(\frac{\dot \Omega}{N}\right)^2+6\frac{\dot
\Omega \dot N}{N^3}$ and $\sqrt{-g}=Ne^{3\Omega}$; then, the~ total
Lagrangian density (\ref{lagrangian}) for the generic-like Universe
is
\begin{equation}
   {\cal L} =  e^{3\Omega} \left[-6 \frac{\ddot \Omega
  }{N}-12\frac{\left( \dot \Omega\right)^2}{N}+6\frac{\dot \Omega \dot N}{N^2}
  - \left(\frac{1}{2}\right)^{\alpha}\left( \dot \phi\right)^{2\alpha} N^{-2\alpha+1}
  \right]. \label{lagran}
\end{equation}

Thus, by~using the total time derivative
$\left(-6e^{3\Omega}\frac{\dot \Omega}{N}
\right)^{\bullet}=-6e^{3\Omega}\frac{\ddot \Omega}{N} -18\,
e^{3\Omega}\frac{\left(\dot \Omega\right)^2}{N}
+6e^{3\Omega}\frac{\dot \Omega \dot N}{N^2}$ in (\ref{lagran}), we
obtain
\begin{equation}
 {\cal L} =  e^{3\Omega} \left[6 \frac{\dot \Omega^2 }{N}
  - \left(\frac{1}{2}\right)^{\alpha}\left( \dot \phi\right)^{2\alpha} N^{-2\alpha+1}
  \right].\label{lala}
\end{equation}

Using the standard definition  of the  momenta $
\Pi_{q^\mu}=\frac{\partial{\cal L}}{\partial{\dot q^\mu}}$, where $
q^{\mu}=(\Omega, \phi)$, we obtain
\begin{eqnarray}
 \Pi_\Omega&=& \frac{12}{N}e^{3\Omega}\dot \Omega, \quad \rightarrow \quad \dot \Omega=\frac{N}{12}e^{-3\Omega}\Pi_\Omega, \label{a} \\
 \Pi_\phi&=&
-\left(\frac{1}{2}\right)^\alpha\frac{2\alpha}{N^{2\alpha-1}}e^{3\Omega}{\dot
\phi}^{2\alpha -1}, \quad \rightarrow \quad \dot
\phi=-N\left[\frac{2^{\alpha-1}}{\alpha}e^{-3\Omega}\Pi_\phi\right]^{\frac{1}{2\alpha
-1}},\label{ph}
\end{eqnarray}
and introducing them into the Lagrangian density, we obtain the
canonical Lagrangian $ {\cal L}_{canonical}=\Pi_{q^\mu} \dot q^\mu
-N {\cal H}$ as
\begin{eqnarray}
 {\cal L}_{canonical}&=& \Pi_{q^\mu} \dot q^\mu
-\frac{N}{24}e^{-\frac{3}{2\alpha-1}\Omega} \left\{
e^{-\frac{6(\alpha-1)}{2\alpha-1}\Omega}\Pi_\Omega^2 -
\frac{12(2\alpha-1)}{\alpha}\,\Pi_\phi^{\frac{2\alpha}{2\alpha-1}}\right\}.
\label{canonical}
\end{eqnarray}
Performing the variation with respect to the lapse function $N$,
${\delta{\mathcal L}}_{canonical}/\delta N=0$, the~Hamiltonian
constraint $\mathcal H=0$ is obtained, where the classical density
is written as
\begin{equation}
 {\cal H}=\frac{1}{24}e^{-\frac{3}{2\alpha-1}\Omega} \left\{
e^{-\frac{6(\alpha-1)}{2\alpha-1}\Omega}\Pi_\Omega^2 -
\frac{12(2\alpha-1)}{\alpha}\left(\frac{2^{\alpha-1}}{\alpha}
\right)^{\frac{1}{2\alpha-1}}\,\Pi_\phi^{\frac{2\alpha}{2\alpha-1}}\right\}.
\label{hami}
\end{equation}

In this point, we return to the equation of the scalar field
(\ref{ss1}) writing $d\tau=Ndt$
\begin{equation}
 \Delta \phi=\sqrt{2}
{\left(\frac{\eta}{\alpha^2}\right)^{\frac{1}{2(2\alpha-1)}}}  \int
e^{-\frac{3}{2\alpha-1}\Omega} Ndt,
\end{equation}
and considering the gauge $ N=24 e^{\frac{3}{2\alpha-1}\Omega}$; the
classical scalar field
 goes like
\begin{equation}
 \phi(t)=\phi_i(t_i)+ 24\sqrt{2}
{\left(\frac{\eta}{\alpha^2}\right)^{\frac{1}{2(2\alpha-1)}}}
(t-t_i), \label{solution}
\end{equation}
where $t_i$ is the initial time for generic epoch and $\phi(t_i)$ is
the scalar field evaluated in this time. In~this way, the~scalar
field is present in the following epochs in our Universe.
However, when we use the equation for momentum (\ref{ph}) in time
$\tau$, we have
$\frac{d\phi}{d\tau}=-\left(\frac{2^{\alpha-1}}{\alpha}e^{-3\Omega}
\Pi_\phi \right)^{\frac{1}{2\alpha-1}}$, and~using the first time
derivative of the scalar field (\ref{phi-tau}), we obtain
$\Pi_\phi^{\frac{1}{2\alpha-1}}=-\sqrt{2}\left(\frac{\eta}{2^{2(\alpha-1)}}\right)^{\frac{1}{2(2\alpha-1)}}$.
Plugging this back into the Hamiltonian constraint, we fiind that
the momenta in the variable $\Omega$ become
$$\Pi_\Omega=2\sqrt{\frac{6(2\alpha-1)}{\alpha}}\,
\left(\frac{2^{\alpha-1}}{\alpha} \right)^{\frac{1}{2(2\alpha-1)}}
\,\left(\frac{\eta}{2^{2(\alpha-1)}}
\right)^{\frac{\alpha}{2(2\alpha-1)}}\,
e^{\frac{3(\alpha-1)}{2\alpha-1}\Omega}.$$

Now, using Equation~(\ref{a}) at time $\tau$, we find that the scale
factor becomes
\begin{equation}
 A(\tau)= \left[
\frac{\alpha}{2(2\alpha-1)}\,\sqrt{\frac{6(2\alpha-1)}{\alpha}}\,\left(\frac{2^{\alpha-1}}{\alpha}
\right)^{\frac{1}{2(2\alpha-1)}}
  \left(\frac{\eta}{2^{2(\alpha-1)}}
\right)^{\frac{\alpha}{2(2\alpha-1)}}
(\tau-\tau_i)\right]^{\frac{2\alpha-1}{3\alpha}},
\label{scale-factor}
\end{equation}
which is consistent with the result obtained in the
Ref.~\cite{berbena}, Equations~(6, 34), in~the time $\tau$,
for~ordinary matter $p =\omega \rho$. When we substitute the
barotropic parameter $\omega=\frac{1}{2\alpha-1}$ in  Equation~(34)
of the paper~\cite{berbena}, we obtain the power law in the time
$\tau$, resulting in the same behavior as in (\ref{scale-factor}).
In this sense, we mention that the kinetic energy of the scalar
field in~the k-essence formalism falsifies the standard~matter.

 In the following, we will place all our effort
in solving the quantum fractionary Wheeler--DeWitt equation.
\section{Quantum~Regime }\label{sec3}
The WDW equation for these models is obtained  by making the usual
substitution $ \Pi_{q^\mu}=-i \hbar \partial_{q^\mu}$ into (\ref
{hami}) and promoting the classical Hamiltonian density in the
differential operator, applied to the wave function
$\Psi(\Omega,\phi)$, $\hat{\cal H}\Psi=0$. Then, we have
\begin{equation}
-\hbar^{2} e^{-\frac{6(\alpha-1)}{2\alpha-1}\Omega}\frac{\partial^2
\Psi}{\partial
\Omega^2}-\frac{12(2\alpha-1)}{\alpha}\,\hbar^{\frac{2\alpha}{2\alpha-1}}\left(\frac{2^{\alpha-1}}{\alpha}
\right)^{\frac{1}{2\alpha-1}}\frac{\partial^{\frac{2\alpha}{2\alpha-1}}}{\partial
\phi^{\frac{2\alpha}{2\alpha-1}}}\Psi=0. \label{q-wdw}
\end{equation}

We noted that the fractional differential equation with degree
$\beta=\frac{2\alpha}{2\alpha-1}$ belongs to different intervals,
depending on the value of the barotropic parameter; that is, when
$\omega_X \in [0,1]$, the~ degree belongs to the interval $[1,2]$,
and when $ \omega_X \in [-1,0)$, the~degree belongs to the interval
$[0,1)$, for~the scalar field $\phi$ (for this calculation, we
remember that $\alpha=\frac{1}{2}\left(1+\frac{1}{\omega_X}
\right)$). It is well-known in standard quantum cosmology that the
best candidates for quantum solutions are those that have a damping
behavior with respect to the scale factor; then, we use this
conjecture in this~formalism.

 For simplicity, the~factor $ e^{-\frac{6(\alpha-1)}{2\alpha-1}\Omega}$ may be the factor
ordered with $ \hat \Pi_\Omega$ in many ways. Hartle and
Hawking~\cite{HH} suggested what  might be called semi-general
factor ordering, which, in~this case, would order the terms  $
e^{-\frac{6(\alpha-1)}{2\alpha-1}\Omega} \hat \Pi^2_\Omega$ as $ -
e^{-( \frac{6(\alpha-1)}{2\alpha-1}- Q)\Omega}\,
\partial_\Omega e^{-Q\Omega}
\partial_\Omega= - e^{-\frac{6(\alpha-1)}{2\alpha-1}\Omega}\, \partial^2_\Omega + Q\,
e^{-\frac{6(\alpha-1)}{2\alpha-1}\Omega} \partial_\Omega, $
where $Q$ is any real constant that measures the ambiguity in the
factor ordering in the variables $ \Omega$  and its corresponding
momenta. We will assume in the following that this factor ordering
for the Wheeler--DeWitt equation becomes
\begin{equation}
-\hbar^{2} e^{-\frac{6(\alpha-1)}{2\alpha-1}\Omega}\frac{\partial^2
\Psi}{\partial \Omega^2}+Q\hbar^2
e^{-\frac{6(\alpha-1)}{2\alpha-1}\Omega}\frac{\partial
\Psi}{\partial \Omega} - \frac{12(2\alpha-1)}{\alpha}
\hbar^{\frac{2\alpha}{2\alpha-1}}\left(\frac{2^{\alpha-1}}{\alpha}
\right)^{\frac{1}{2\alpha-1}}\frac{\partial^{\frac{2\alpha}{2\alpha-1}
} }{\partial \phi^{\frac{2\alpha}{2\alpha-1}}}\Psi=0, \label{wdw}
\end{equation}
which when written in terms of the  $\beta$ parameter, becomes
\begin{equation}
-\hbar^{2} e^{-3(2-\beta)\Omega}\frac{\partial^2 \Psi}{\partial
\Omega^2}+Q\hbar^2 e^{-3(2-\beta)\Omega}\frac{\partial
\Psi}{\partial \Omega} -
\frac{24}{\beta}\left(\frac{2^{\alpha-1}}{\alpha}
\right)^{\frac{1}{2\alpha-1}} \hbar^{\beta}\frac{\partial^\beta
}{\partial \phi^\beta} \Psi=0. \label{wdw}
\end{equation}

By employing the separation variables method for the wave function
$\Psi={\cal A}(\Omega)\, {\cal B}(\phi)$, we have the following two
differential equations for $(\Omega,\phi)$
\begin{eqnarray}
 \frac{d^2 {\cal A}}{d\Omega^2}\, - Q \frac{d {\cal A}}{d \Omega}
\mp \frac{\mu^2}{\hbar^2}e^{3(2-\beta)\Omega}{\cal A}
 &=& 0,\label{scale}\\
 \frac{d^\beta {\cal B_\pm}}{ d \phi^\beta} \pm
\left(\frac{\alpha}{2^{\alpha-1}} \right)^{\frac{1}{2\alpha-1}}
\frac{\mu^2\,\beta}{24 \hbar^\beta} {\cal B}_\pm&=&0,\label {phi-1}
\end{eqnarray}
where $ {\cal B}_\pm$ considers the sign in the differential
equation. The~fractional differential Equation~(\ref{phi-1}) can be
given in the fractional frameworks,
following~\cite{Rosales1,Rosales2} and identifying $
\gamma=\frac{\beta}{2}=\frac{\alpha}{2\alpha-1},$ where now, $
\gamma$ is the order of the fractional derivative taking values in $
0<\gamma\leq 1$; then, we can write
\begin{equation}
  \frac{d^{2\gamma} {\cal B}_\pm}{d \phi^{2\gamma}} \pm
\left(\frac{\alpha}{2^{\alpha-1}}
\right)^{\frac{1}{2\alpha-1}}\frac{\gamma \mu^2 }{12
\hbar^{2\gamma}} {\cal B}_\pm=0 ,\qquad 0<\gamma \leq 1,
\label{phi433}
\end{equation}
the solution of the Equation~(\ref{phi433}) with a positive sign may
be obtained by applying direct and inverse Laplace
transforms~\cite{Rosales2}, providing
\begin{equation}
 {\cal B}_+ (\phi, \gamma) =
 \mathbb{E}_{2\gamma}\left(- z^2\right),
 \qquad z=\left(\frac{\alpha}{2^{\alpha-1}}
\right)^{\frac{1}{2(2\alpha-1)}}\frac{\sqrt{\gamma} \mu
}{2\sqrt{3}\hbar^{\gamma}} \phi^\gamma,\qquad 0<\gamma \leq 1.
\label{R}
 \end{equation}

In the ordinary case, $\gamma=1$; then, the~solution
is~\cite{Rosales2},
\begin{equation}
{\cal B}_+ (\phi, 1) =\mathbb{E}_2\left[-
\left(\frac{\mu}{2\sqrt{3}\hbar} (\phi-\phi_0) \right)^2\right]=
cos\left(\frac{\mu}{2\sqrt{3}\hbar} (\phi-\phi_0) \right),
\label{clasico1}
\end{equation}
which is in agreement with the Equation~(\ref{cases}),  employing
the Taylor~series.

 Following the book of Polyanin~\cite{polyanin} (page 179.10), we discovered
the solution for the first equation, considering different values in
the factor ordering parameter (we take the corresponding sign minus
in the constant $\mu^2$)
\begin{equation}  {\cal A}= A_0\, e^{\frac{
Q\Omega}{2}}\,Z_\nu\left[\frac{2\mu}{3\hbar(2-\beta)}\sqrt{- 1}
e^{\frac{3(2-\beta)}{2}\Omega} \right]=A_0\, e^{\frac{
Q\Omega}{2}}\,K_\nu\left[\frac{\mu}{3\hbar(1-\gamma)}
e^{3(1-\gamma)\Omega} \right],
\end{equation}
with order $  \nu=\pm \frac{Q}{6(1-\gamma)}$, where we had written
the second expression in terms of the fractional order
$\gamma=\frac{\beta}{2}$, and~the solutions which become dependent
on the sign of its argument; when $\sqrt{1}$ (for ${\cal B}_-$),
the~Bessel function $Z_\nu$ becomes the ordinary Bessel function
$J_\nu$. When $\sqrt{-1}$ (for ${\cal B}_+$), this becomes the
modified Bessel function $K_\nu$. For~the particular values
$\beta=2$ ($\gamma=1$), it will be necessary to solve the original
differential equation for this~variable.

Then, we have the probability density  $ |\Psi|^2$ by considering
only $ {\cal B}_+$, $\gamma \not= 1$,
\begin{equation}
 |\Psi|^2=\psi_0^2 \,e^{Q\Omega}\, \mathbb{E}^2_{2\gamma}\left(-
z^2\right)\,\,K_\nu\left[\frac{\mu}{3\hbar(1-\gamma)}
e^{3(1-\gamma)\Omega} \right], \quad
z=\left(\frac{\alpha}{2^{\alpha-1}}
\right)^{\frac{1}{2(2\alpha-1)}}\frac{\sqrt{\gamma} \mu
}{2\sqrt{3}\hbar^{\gamma}} \phi^\gamma. \label{dens}
\end{equation}

On the other hand, it is well-known that in standard quantum
cosmology, the wave function is unnormalized. There is no systematic
method to do this, as the Hamiltonian density is not Hermitian.
In~particular cases, wave packets can be constructed, and  from
these wave packets  we can construct a normalized wave function.
In~this work, we could not construct these wave packets. We hope to
be able to do it in future~studies.

 In the following, we present particular cases in the evolution
of the Universe and some plots by employing the
Equation~(\ref{dens}), and~for better viewing in the plots, we
introduce by hand particular values to the constant $\psi_0$.
\begin{enumerate}
\item{} Radiation epoch, $\omega_X=\frac{1}{3}$, $\alpha=2, \to \beta=\frac{4}{3} \to \gamma=\frac{2}{3}$ .

 When we choose the radiation case, (\ref{dens}) is written as
\begin{equation}
 |\Psi|^2=\psi_0^2 \,e^{Q\Omega}\,
\mathbb{E}^2_{\frac{4}{3}}\left(-
z^2\right)\,\,K_{\frac{Q}{2}}^2\left[\frac{\mu}{\hbar} e^{\Omega}
\right], \quad
z=\frac{\mu}{3\sqrt{2}\hbar^{\frac{2}{3}}}\phi^{\frac{2}{3}}.
\label{radiation}
\end{equation}

In the following Figure~\ref{radiation-era}, we take  the
probability density (\ref{radiation}); in the first and second
Figures, and~for better viewing in the plots, we take the constant
$\psi_0=\frac{1}{10}$, and~in the third Figure the value becomes 1.
In all Figures, the~behavior of the probability density, in~both
variables ($\Omega,\phi$), has the appropriate decadent behavior.
The range of the variable equals to $\phi \in [0,3000], [0,200]$,
and $[0,40]$, respectively.

\begin{figure}[h]
\begin{tabular}{cc}
\includegraphics[totalheight=0.2\textheight]{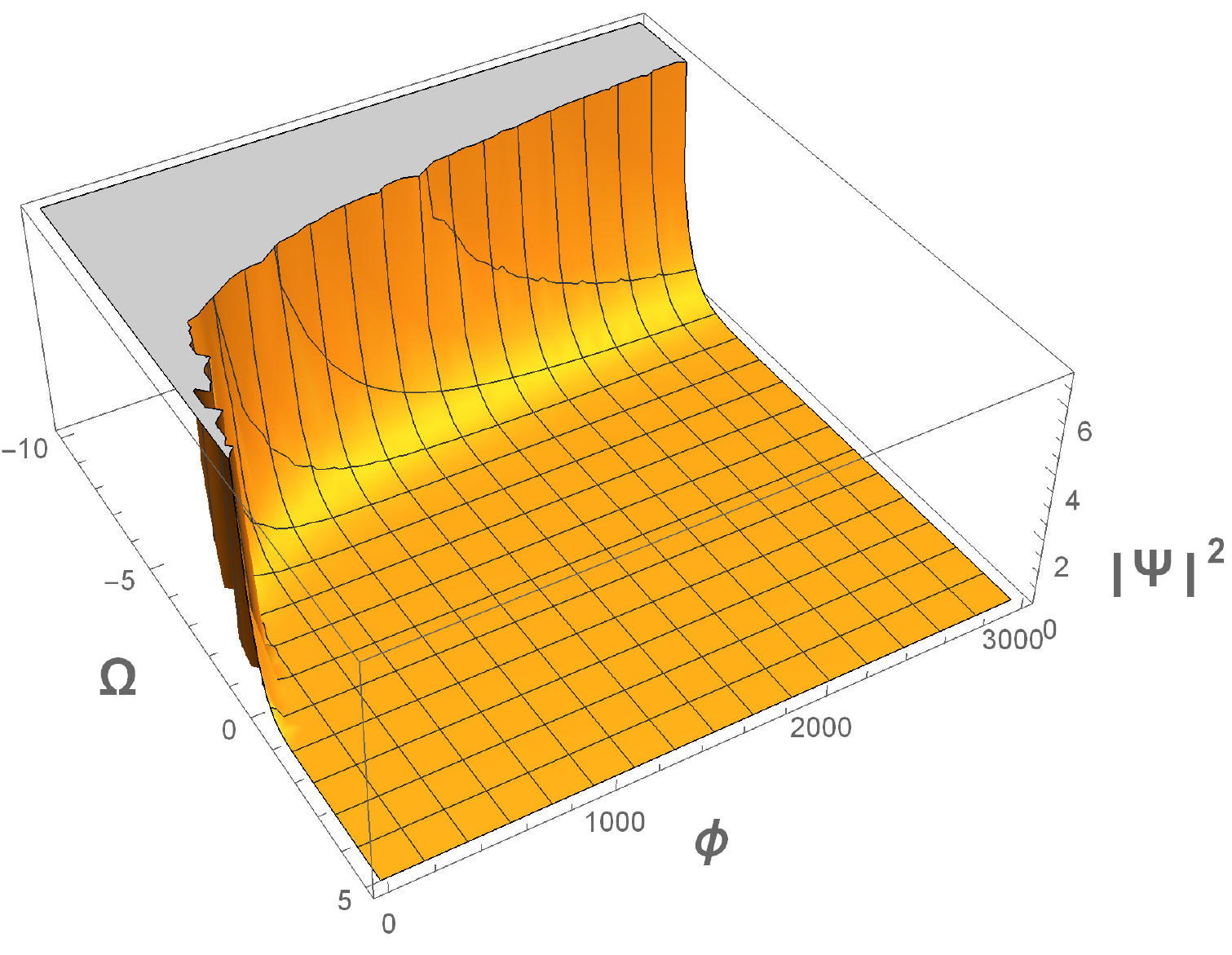}
&\includegraphics[totalheight=0.17\textheight]{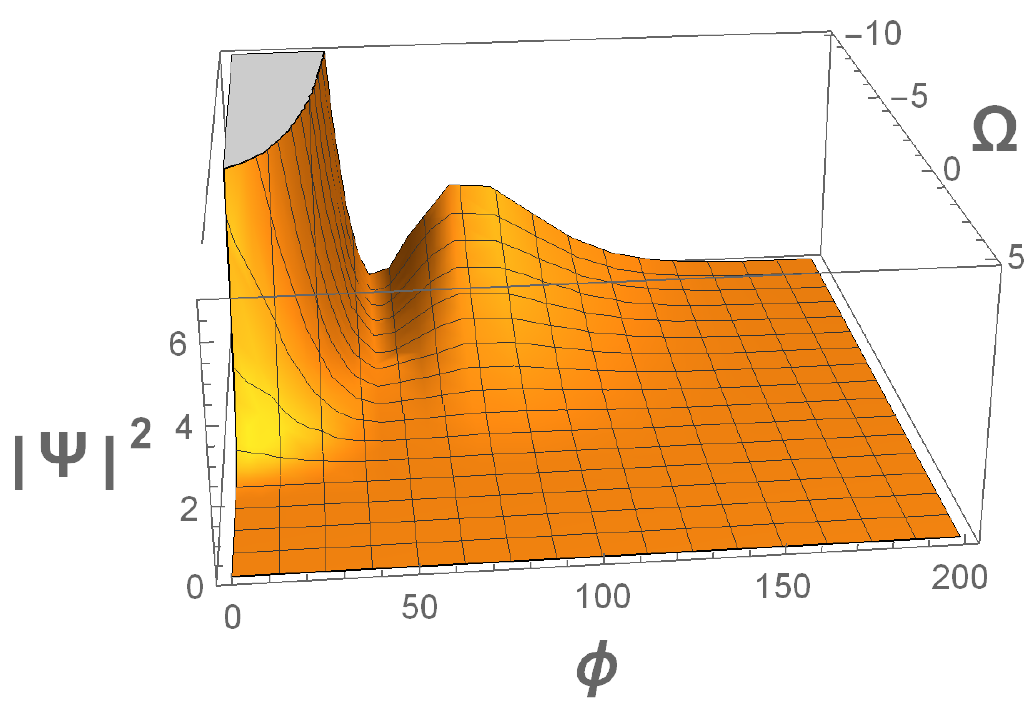} \\
\multicolumn{2}{c}{\includegraphics[totalheight=0.2\textheight]{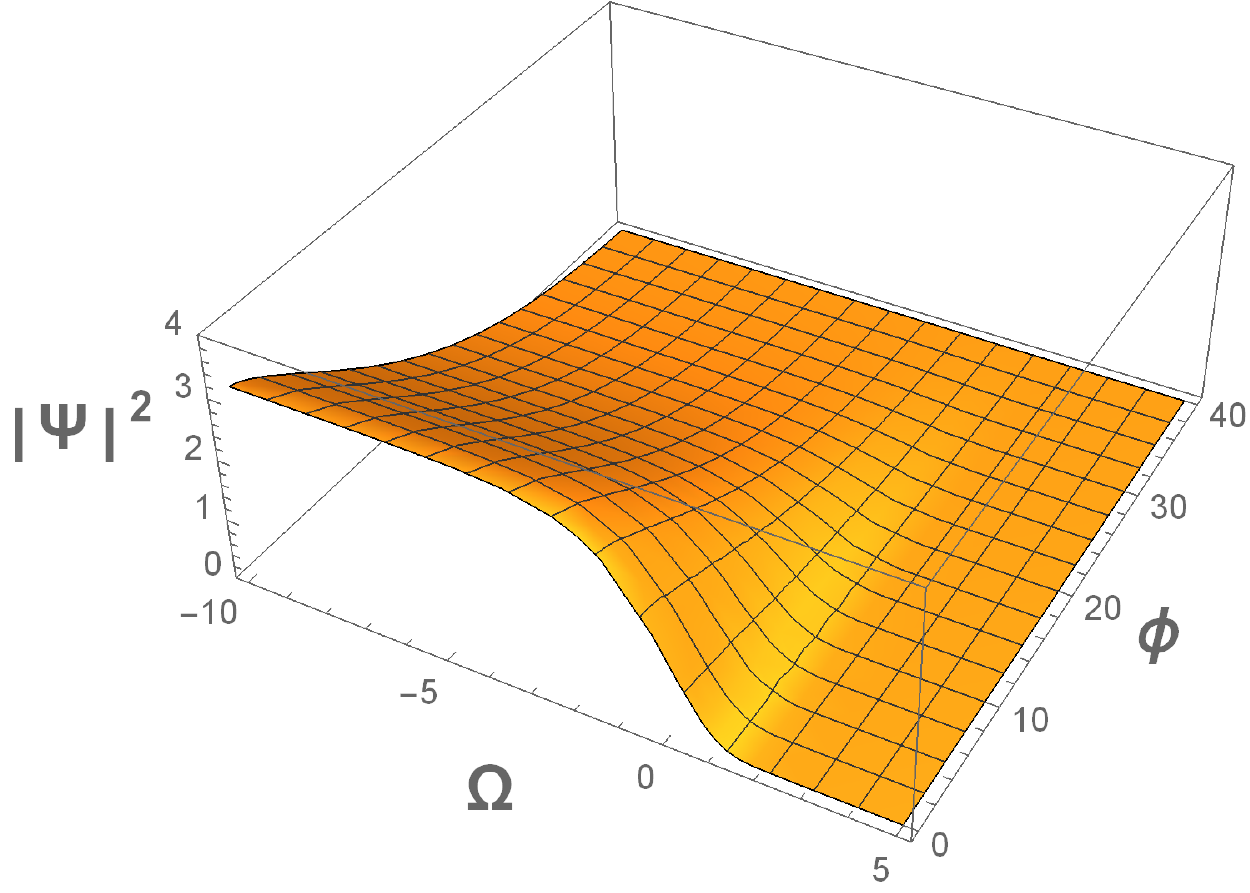} } \\
\end{tabular}
\caption{In the radiation era, we plot the
 Equation~(\ref{radiation}), considering different values
in the order parameter Q $=-1,1,0$, from top to bottom,
respectively. We consider the value for the parameter $ \mu=0.5$; we
discard other values in the Q parameter.} \label{radiation-era}
\end{figure}

\item{}  Solution to $\omega_X=\frac{2}{3}$, $\alpha=\frac{5}{4}, \to \beta=\frac{5}{3} \to \gamma=\frac{5}{6}$.

 The probability density of the wave function becomes (here,
$z=\left(\frac{5}{4\sqrt[4]{2}}
\right)^{\frac{1}{3}}\frac{\sqrt{5}\mu}{6\sqrt{2}\hbar^{\frac{5}{6}}}\phi^{\frac{5}{6}})$
\begin{equation}
 \Psi^2=  \psi_0^2\, e^{Q\Omega} \mathbb{E}_{\frac{5}{3}}^2(-
z^2)\,K^2_Q\left[\frac{\mu}{\hbar} e^{\frac{1}{2}\Omega} \right].
\label{d-radiation-like}
\end{equation}
In the  Figure~\ref{radiation-like}, we take  the probability
density (\ref{d-radiation-like}); in the first and second Figures,
and for better viewing in the plots, we take the constant
$\psi_0=\frac{1}{\sqrt{10}}$, and~in the third Figure the value
becomes 1. In~all Figures, the~behavior of the probability density,
in both variables ($\Omega,\phi$), has the appropriate decadent
behavior, and~it presents an oscillatory behavior when $\omega_X \to
1$, since that is the behavior according to the
Equation~(\ref{clasico1}).  Only for $Q=-1$, the~probability density
has a moderate increase in the direction where the scalar
field~evolves.

\item{} Dust era, $ \omega_X=0$, $\alpha\to \infty$; thus, $\beta =1 \to \gamma=\frac{1}{2}$.
In the dust case, the~solution  for the scale factor becomes
\begin{equation}  {\cal A}= A_0\, e^{\frac{
Q\Omega}{2}}\,Z_\nu\left[\frac{\mu}{\hbar}\sqrt{\pm 1}
e^{\frac{3}{2}\Omega} \right], \qquad \nu=\pm \frac{Q}{2}.
\end{equation}
In this case, the~fractional differential Equation~(\ref{phi-1}) for
the scalar field is reduced to the first-order differential equation
(for both signs in $\mu^2$)
$$ \frac{d {\cal B_\mp}}{ d
\phi} \mp \left(\frac{\alpha}{2^{\alpha-1}}
\right)^{\frac{1}{2\alpha-1}}\frac{\mu^2\,}{24 \hbar} {\cal
B_\mp}=0, \qquad \to \qquad {\cal B}_\mp=\beta_\mp \, e^{\pm
\left(\frac{\alpha}{2^{\alpha-1}}
\right)^{\frac{1}{(2\alpha-1)}}\frac{\mu^2}{24\hbar}\Delta \phi}.$$
Then, the~probability density  of the wave function becomes
\begin{equation}
 \Psi^2= \psi_0^2\, \left\{
\begin{tabular}{l}
$ e^{(Q\Omega + \left(\frac{\alpha}{2^{\alpha-1}}
\right)^{\frac{1}{(2\alpha-1)}}\frac{\mu^2}{12\hbar}\Delta
\phi)}\,J^2_{\frac{Q}{3}}\left[\frac{\mu}{\hbar}
e^{\frac{3}{2}\Omega} \right]$\\
$  e^{(Q\Omega - \left(\frac{\alpha}{2^{\alpha-1}}
\right)^{\frac{1}{(2\alpha-1)}}\frac{\mu^2}{12\hbar}\Delta
\phi)}\,K^2_{\frac{Q}{3}}\left[\frac{\mu}{\hbar}
e^{\frac{3}{2}\Omega} \right]$
\end{tabular}
\right. \label{densi}
\end{equation}
In the following Figure~\ref{dust-era}, we present the behavior of
the probability density $\Psi^2$ by using the Equation~(\ref{densi})
and taking the values for the order parameter $ Q=-1,0,1$,
because~with these values, the~probability density presents a
structure well-defined for this era. In~some of them, one structure
did not appear; thus, we gave it a profile for the probability
density for particular values in the scalar field. In~these cases,
the~behavior of our Universe is quite selective in this formalism.
Additionally, we can notice that the probability density has a
moderate increase in the direction where the scalar field evolves.
Similar results were reported in other
formalisms~\cite{chiral,quintom1,quintom2}.

\begin{figure}[h]
\begin{tabular}{cc}
\includegraphics[totalheight=0.15\textheight]{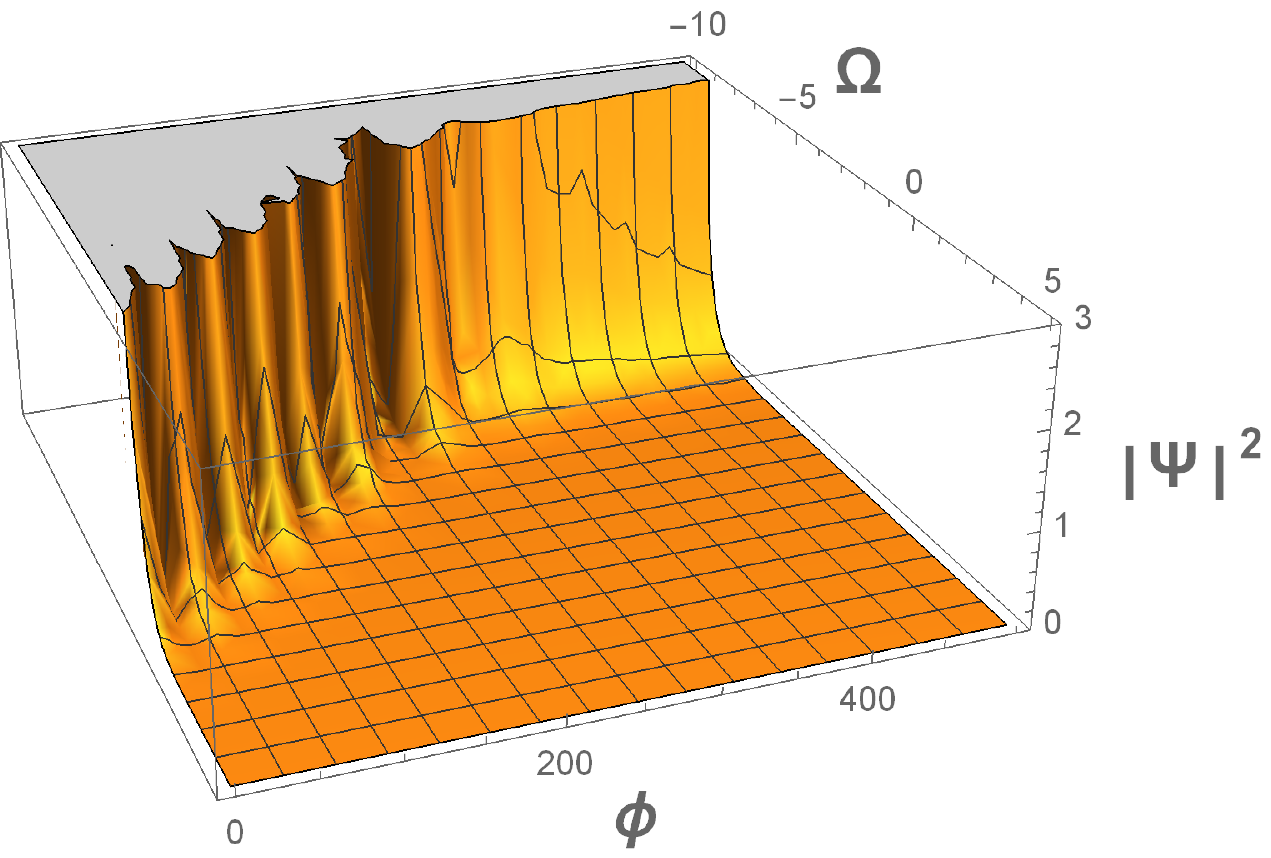} &
\includegraphics[totalheight=0.15\textheight]{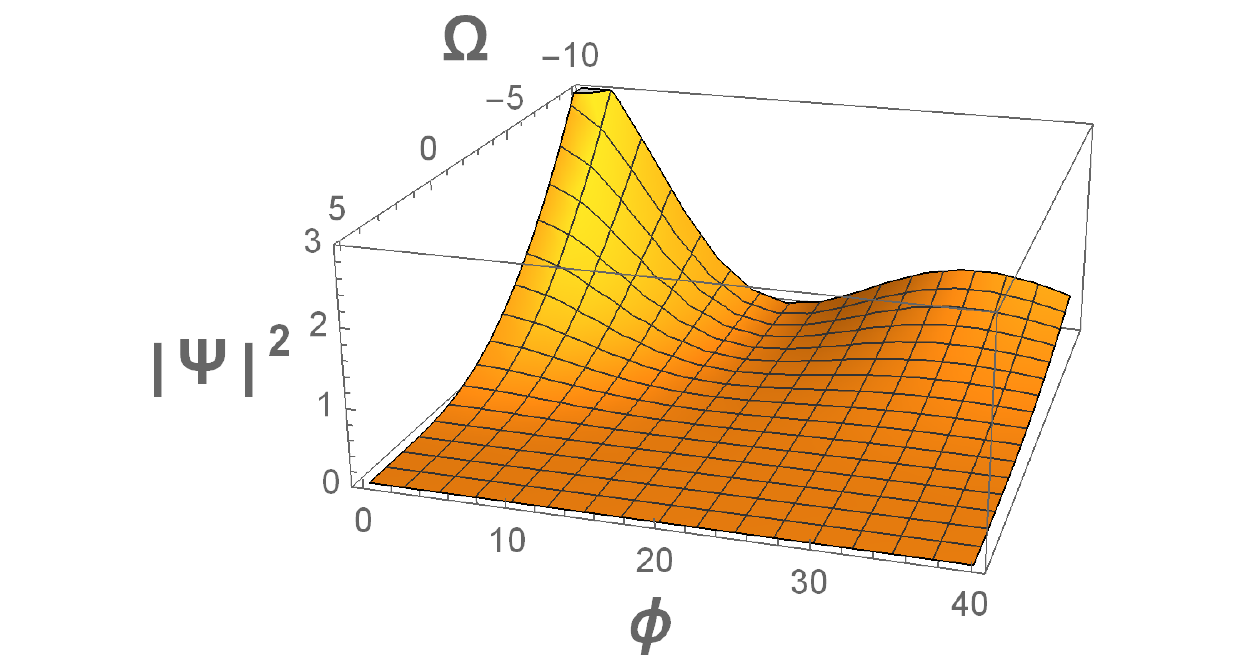} \\
\multicolumn{2}{c}{\includegraphics[totalheight=0.15\textheight]{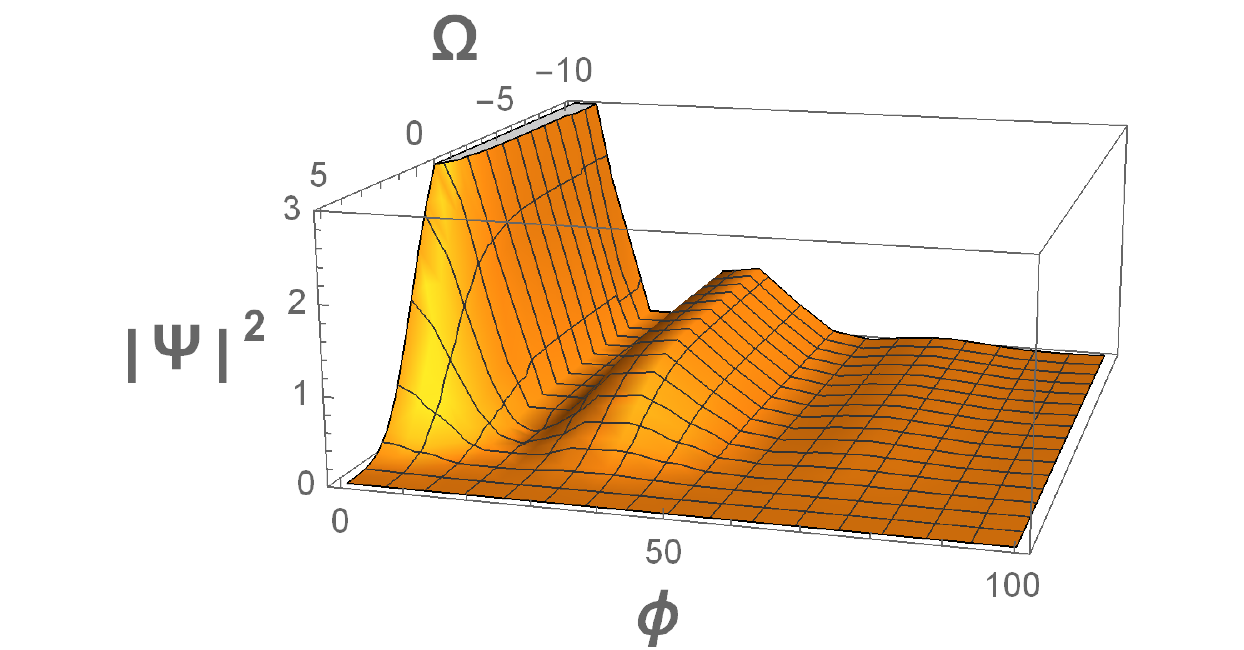}} \\
\end{tabular}
\caption{In the radiation-like era, we plot different combinations
of the Equation~(\ref{d-radiation-like}), considering different
values in the order parameter Q $=-1,0,1$, from top to bottom,
respectively. We consider the value for the parameter $ \mu=0.5$; we
discard other values in the Q parameter.} \label{radiation-like}
\end{figure}

\item{} inflation such as  $ \omega_X=-\frac{1}{3}$, $\alpha=-1$; thus, $\beta =\frac{2}{3} \to \gamma=\frac{1}{3}$.

For this particular case, (\ref{dens}) is written as
\begin{equation}
 \Psi^2=  \psi_0^2\, e^{Q\Omega} \mathbb{E}_{\frac{2}{3}}^2(-
z^2)\,\,K^2_{\frac{Q}{4}}\left[\frac{\mu}{\hbar} 2e^{\Omega}
\right], \label{densi-infla1};
\end{equation}
however, the~argument in the Mittag--Leffer function is complex,
being
\begin{equation}
z=\left(0.6873648184993014 - 0.39685026299204984
I\right)\frac{\mu}{6\hbar^{\frac{1}{3}}}\phi^{\frac{1}{3}},
\label{inf-1-3}
\end{equation}
and the corresponding graph of the probability density can be made
based on this function, taking the Re[z] or Im[z] parts.

\begin{figure}[h]
\begin{tabular}{cc}
\includegraphics[totalheight=0.2\textheight]{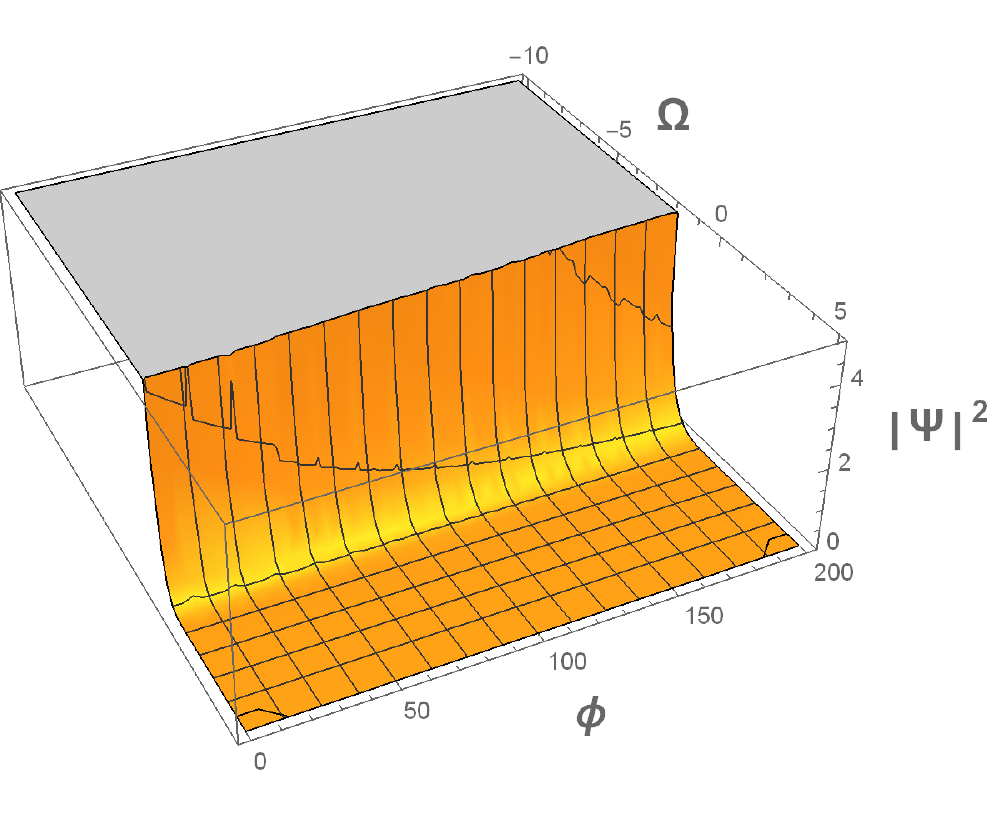}&
\includegraphics[totalheight=0.2\textheight]{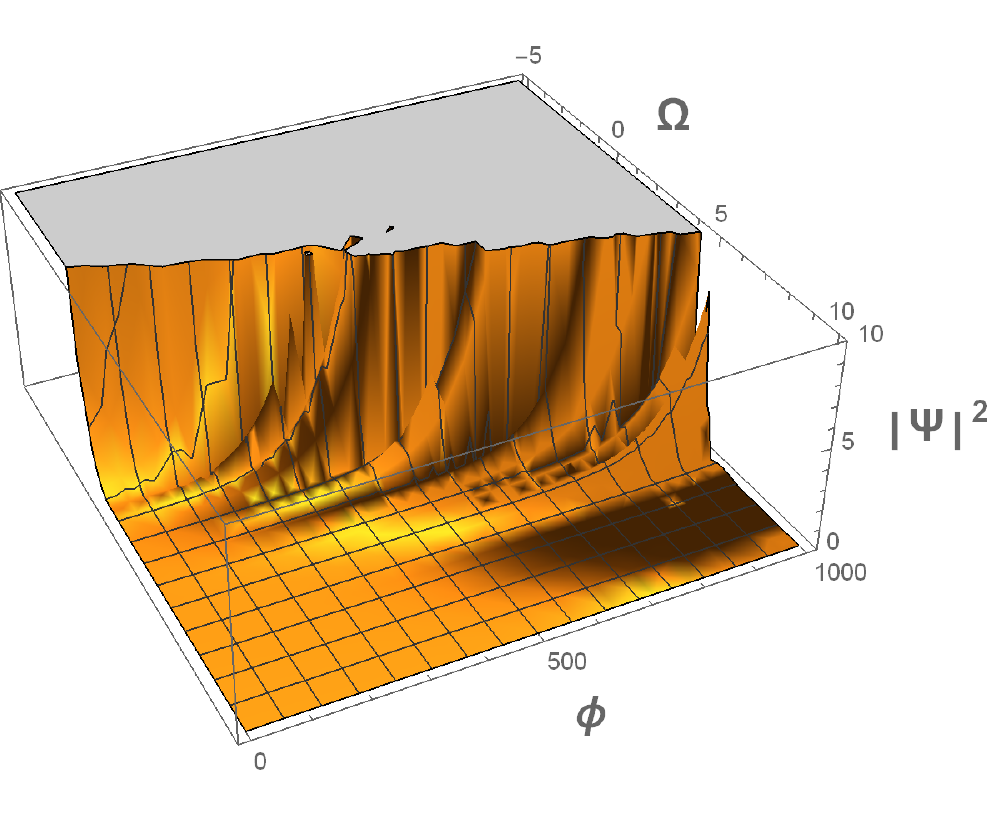} \\
\includegraphics[totalheight=0.2\textheight]{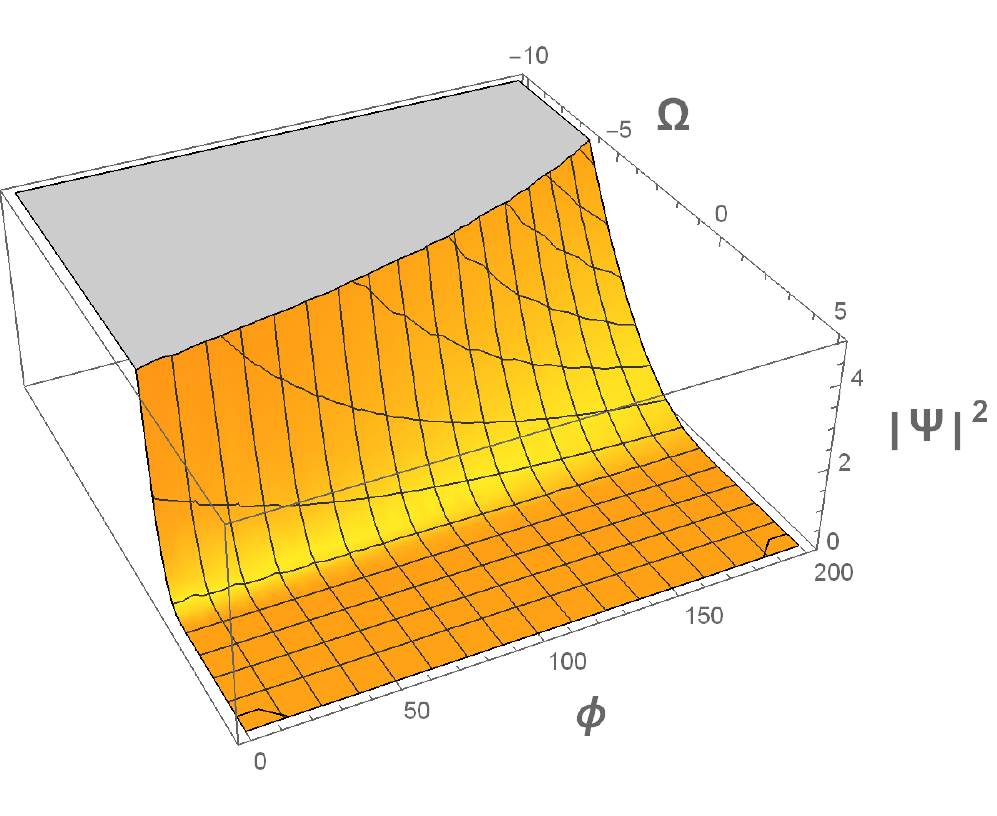} &
\includegraphics[totalheight=0.2\textheight]{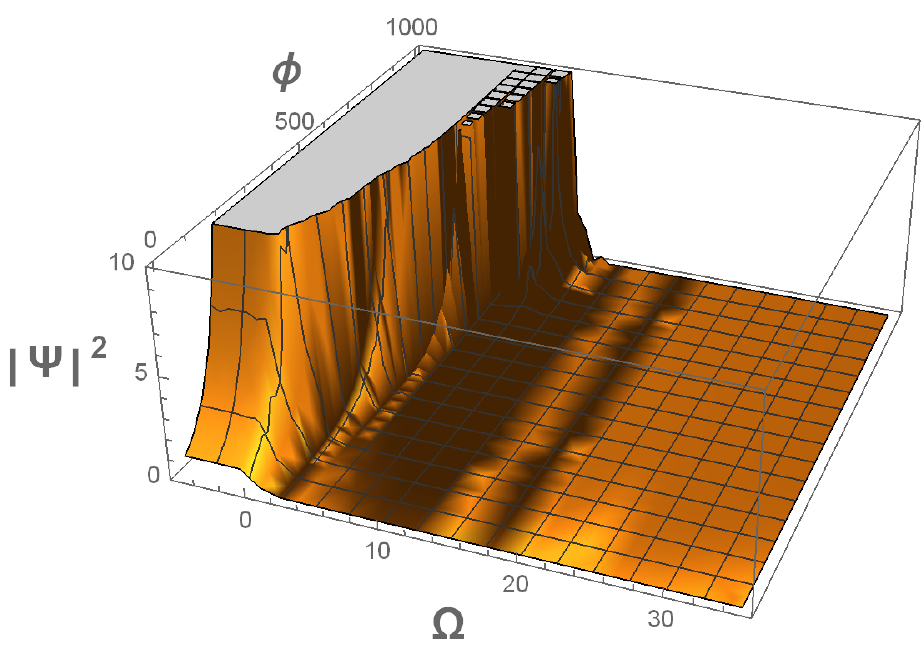} \\
\includegraphics[totalheight=0.2\textheight]{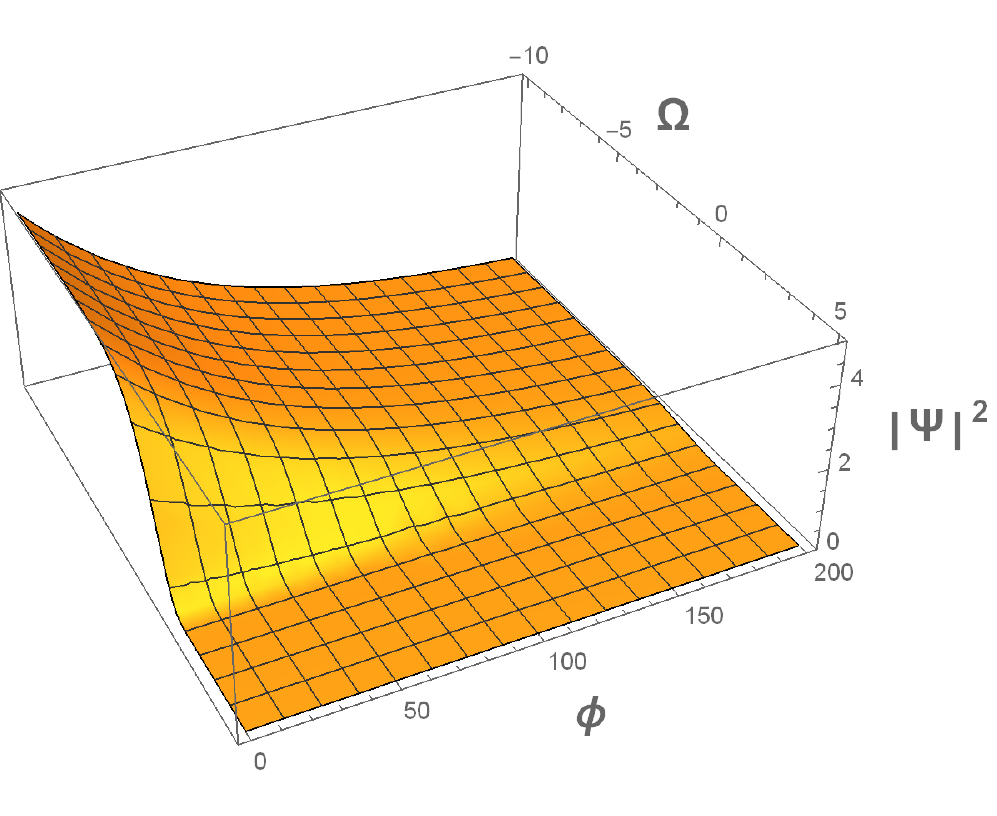} &
\includegraphics[totalheight=0.2\textheight]{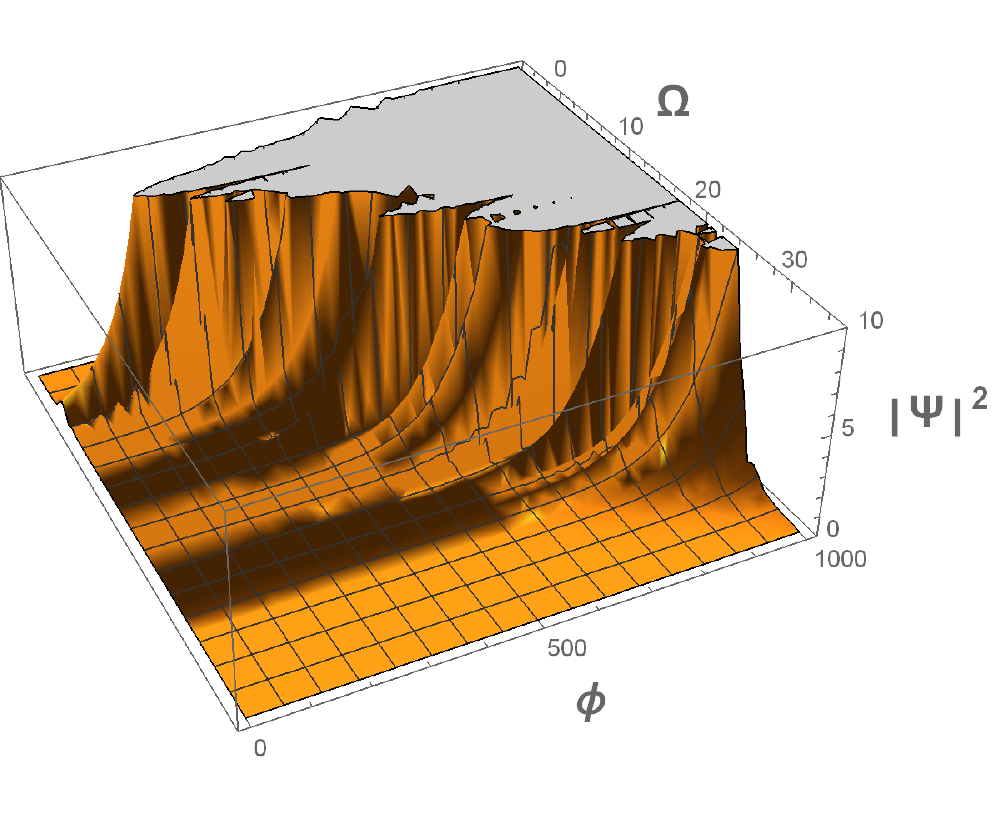} \\
\end{tabular}
\caption{In the dust era, we have the corresponding solution ${\cal
B}_+$ with the modified Bessel function $K_\nu$ and ${\cal B}_-$,
the ordinary Bessel function  of the Equation~(\ref{densi}),
considering different values in the order parameter Q $=-1,0,1$,
from top to bottom, respectively. We consider the value for the
parameter $ \mu=0.5$; we discard other values in the Q parameter.}
\label{dust-era}
\end{figure}

\item{} inflation such as  $ \omega_X=-\frac{2}{3}$, $\alpha=-\frac{1}{4}$; thus, $\beta =\frac{1}{3}\to \gamma=\frac{1}{6}$.

For this particular case, (\ref{dens}) is written as
\begin{equation}
 \Psi^2=  \psi_0^2\, e^{Q\Omega}
\,\mathbb{E}_{\frac{1}{3}}^2(-
z^2)\,\,K^2_{\frac{Q}{5}}\left[\frac{\mu}{\hbar}
e^{\frac{5}{2}\Omega} \right]. \label{densi-infla}
\end{equation}
and the argument of the Mittag--Leffer function, in~this case, is
equal to the previous case, complex
\begin{equation}
z=\left(0.5946035575013603 - 1.0298835719535588
I\right)\frac{\mu}{6\hbar^{\frac{1}{6}}}\phi^{\frac{1}{6}}.\label{z2-3}
\end{equation}
 In a general way, the behavior of the probability density for
both inflation-like scenarios is similar, in~the Re[z] or Im[z]
parts, over~a wide range of values in the scalar field, as~it
appears in Figure~\ref{inflation2}. For the behavior for both
inflation-like cases in the value of $\Omega$, the~behavior is
appropriate.

\end{enumerate}

\begin{figure}[h]
\begin{tabular}{cc}
\includegraphics[totalheight=0.17\textheight]{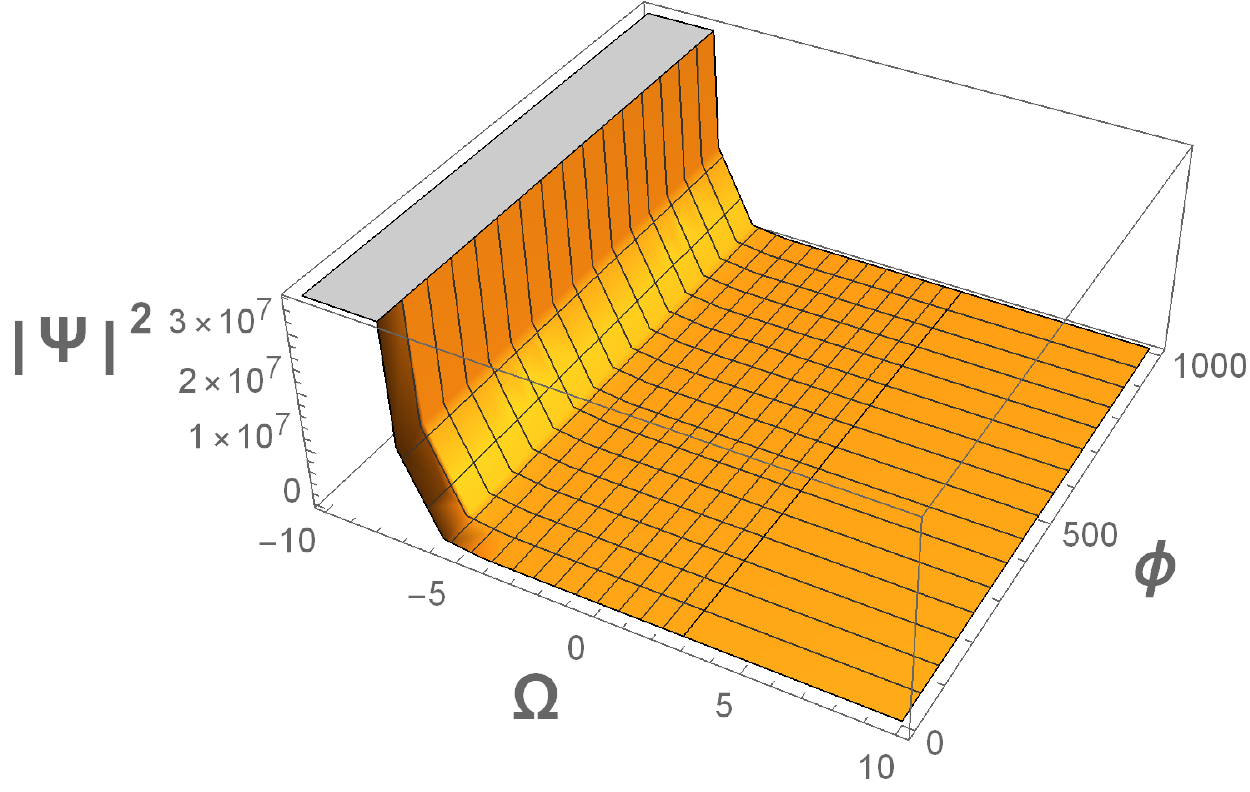}&
\includegraphics[totalheight=0.17\textheight]{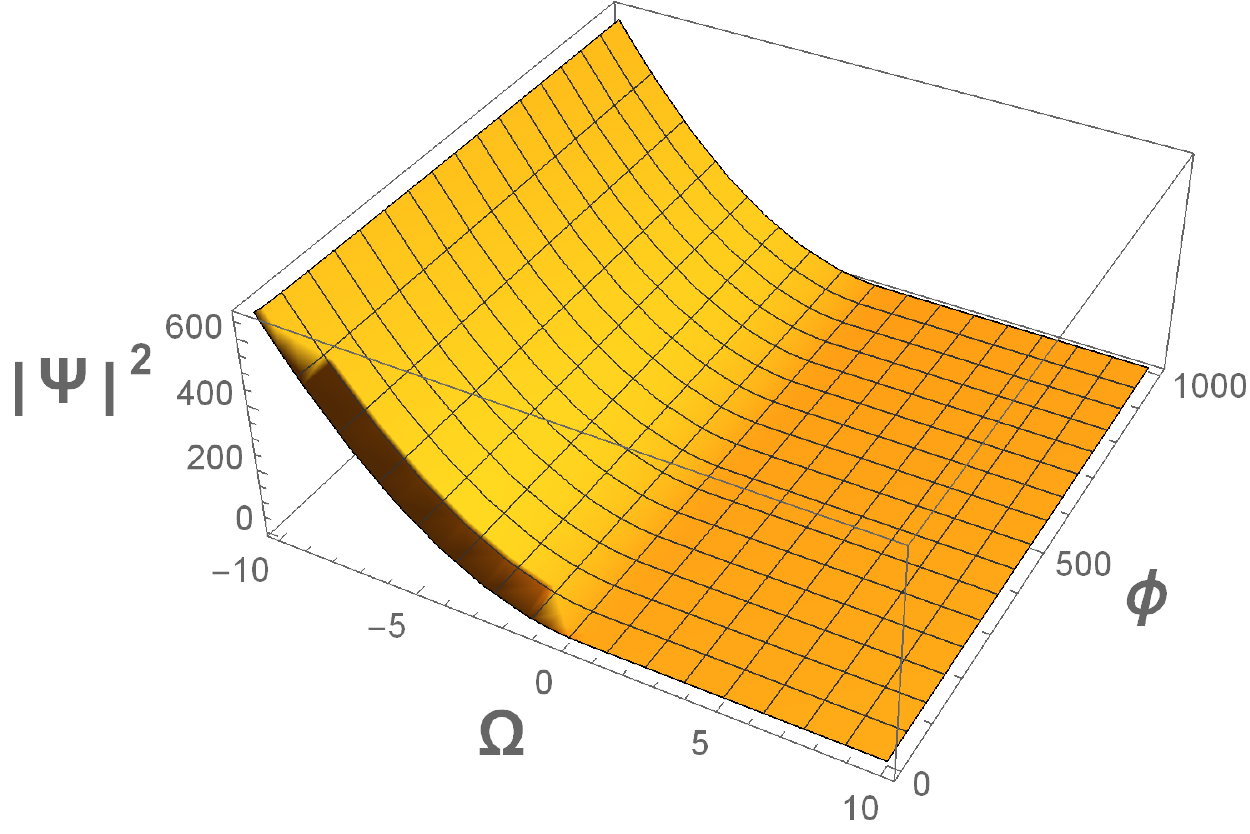} \\
\multicolumn{2}{c}{\includegraphics[totalheight=0.17\textheight]{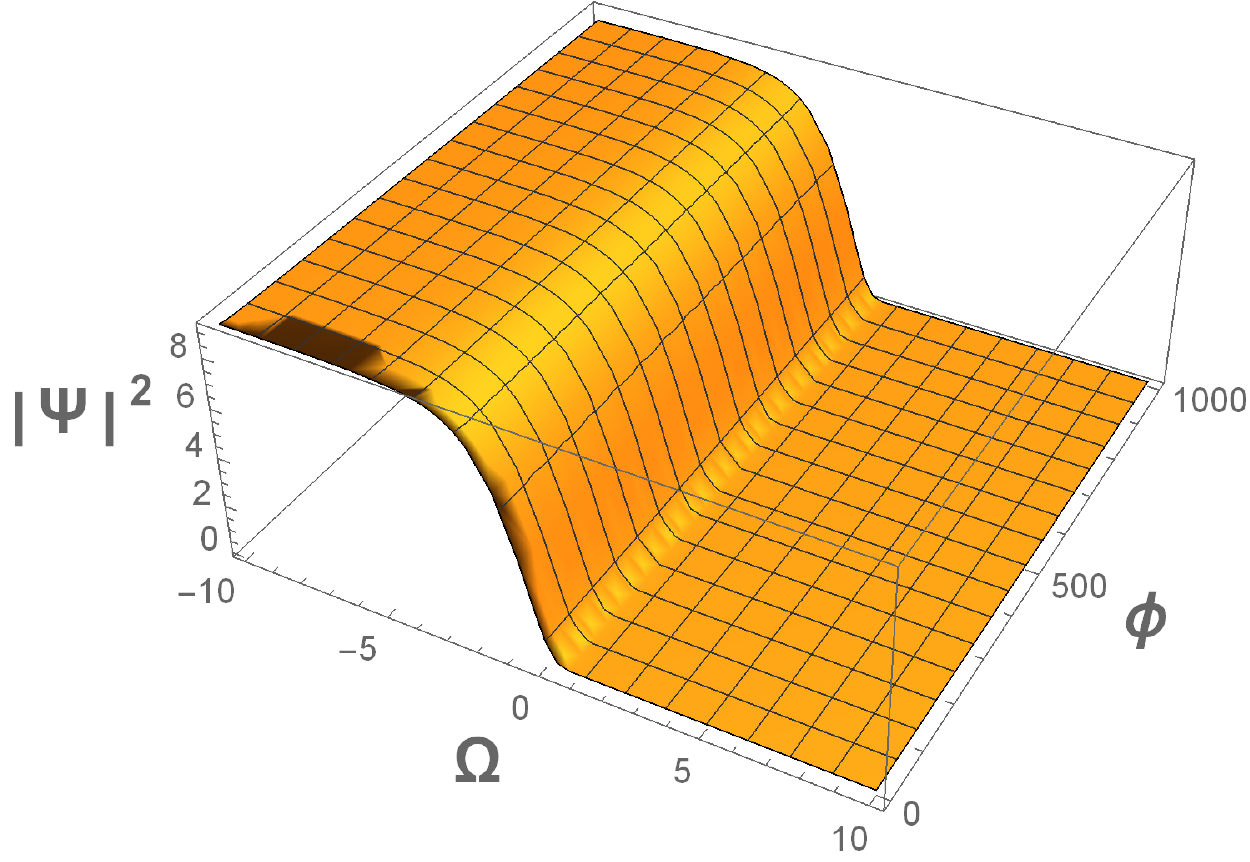}} \\
\end{tabular}
\caption{In the second case of the inflation-like scenario,
corresponding to Equation~(\ref{densi-infla}), the behavior for the
probability density in terms of the Mittag--Leffler function with
complex values in its argument z is shown, for Q $=-1,0,1$. From top
to bottom, respectively, we consider the value for the parameter $
\mu=0.5$. In these graphs, we use the Re[z] only; however, the plots
with  Im[z] are similar.} \label{inflation2}
\end{figure}

\section{Final~Remarks}\label{sec4}
There are different formalisms to incorporate fractional derivatives
to cosmology. One of them starts from the variational principle of
the action of general relativity with a fractional kernel; another
way is starting from a particular configuration, for~example,
the~FLRW model, and changing the ordinary derivatives with
fractional one \cite{Roberts,co1,co2,co3,co4,co5,co6}. Unlike the
previous formalisms, in~the present work, we employed an action that
contains a Lagrangian and a fractional parameter; in~this way,
equations of non-integer order in cosmology are~obtained.

\begin{figure}[h]
\includegraphics[totalheight=0.13\textheight]{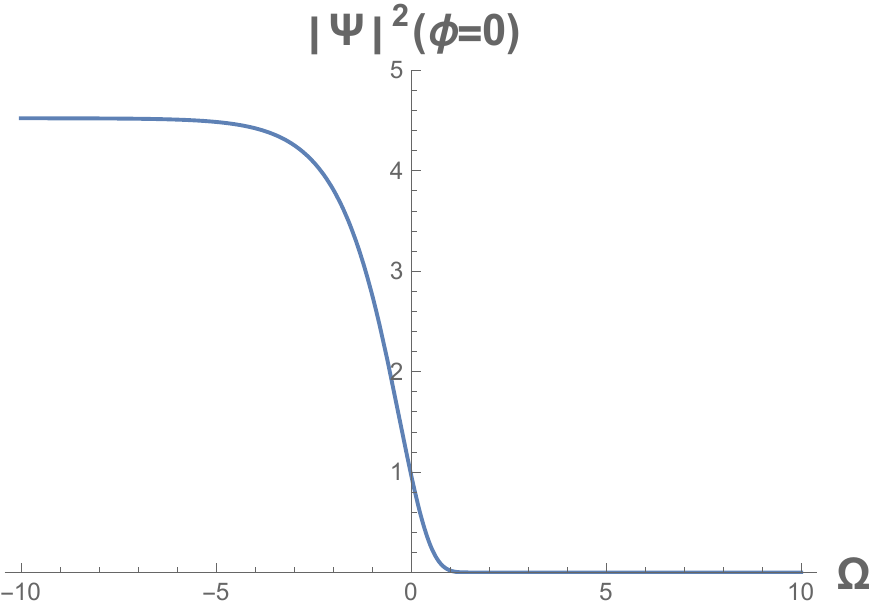}
\includegraphics[totalheight=0.13\textheight]{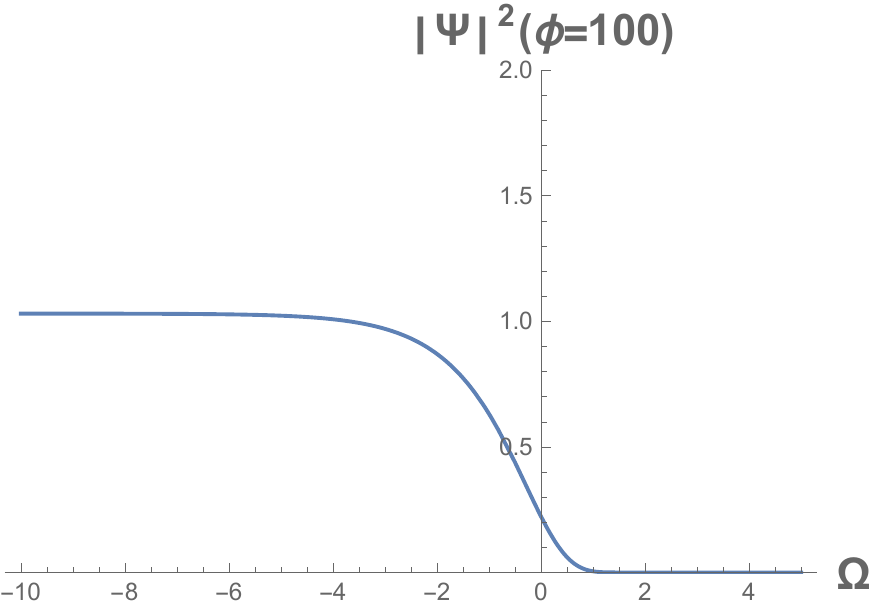}
\includegraphics[totalheight=0.13\textheight]{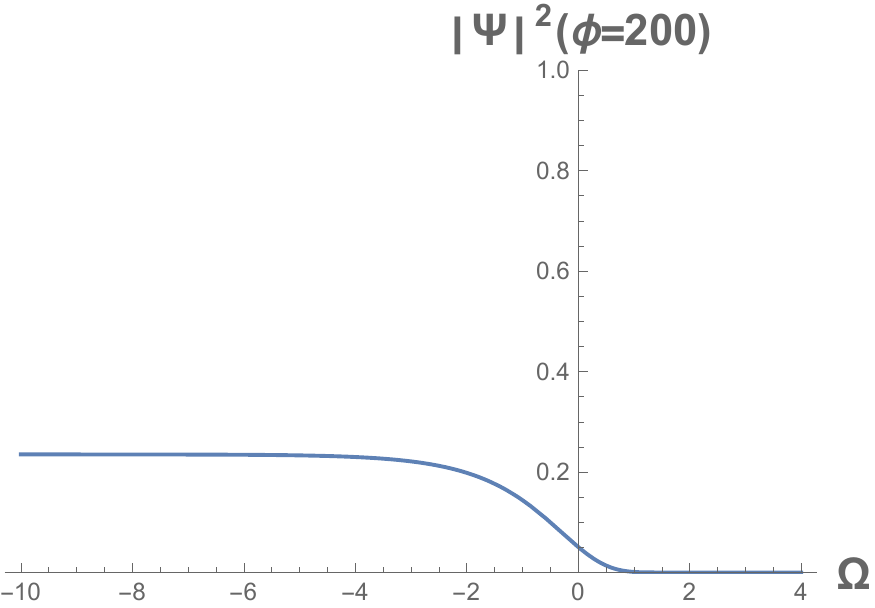}
\includegraphics[totalheight=0.13\textheight]{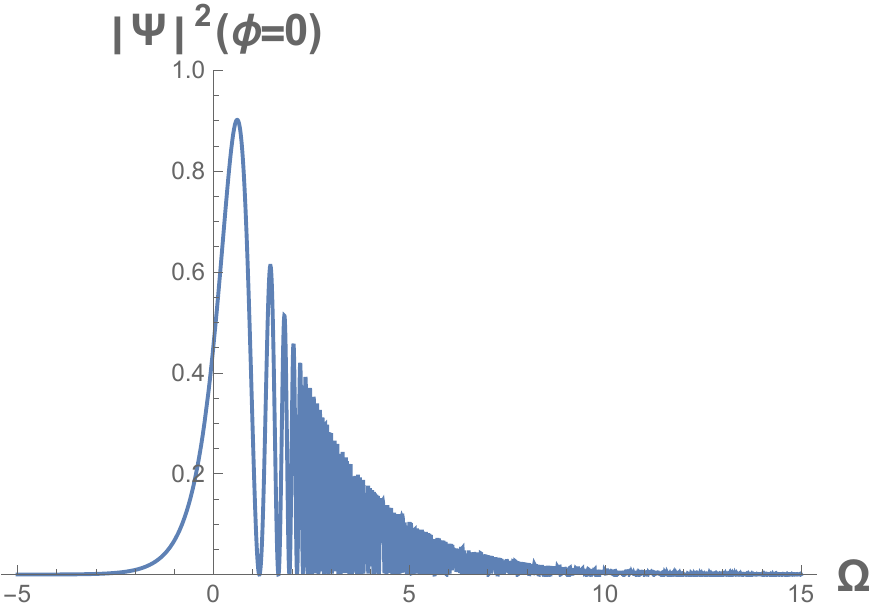}
\includegraphics[totalheight=0.13\textheight]{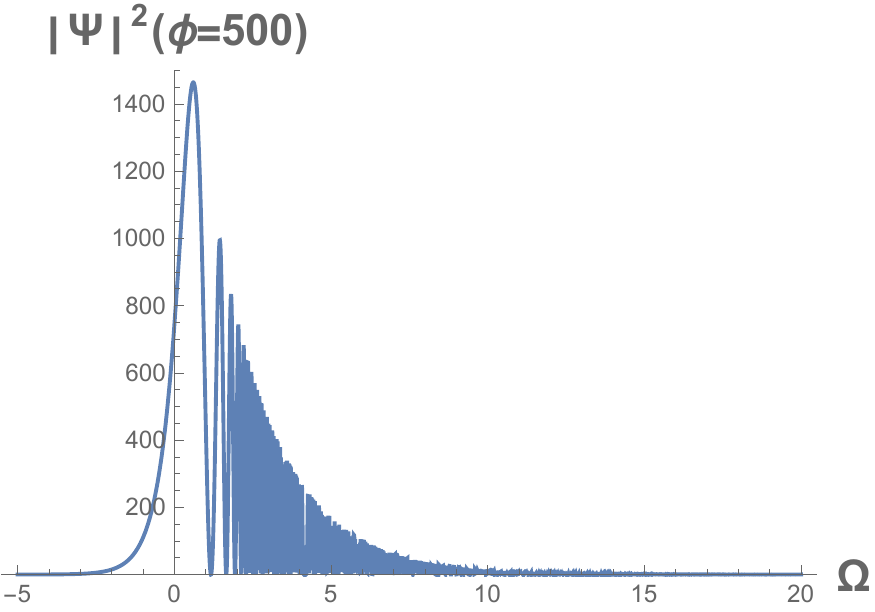}
\includegraphics[totalheight=0.13\textheight]{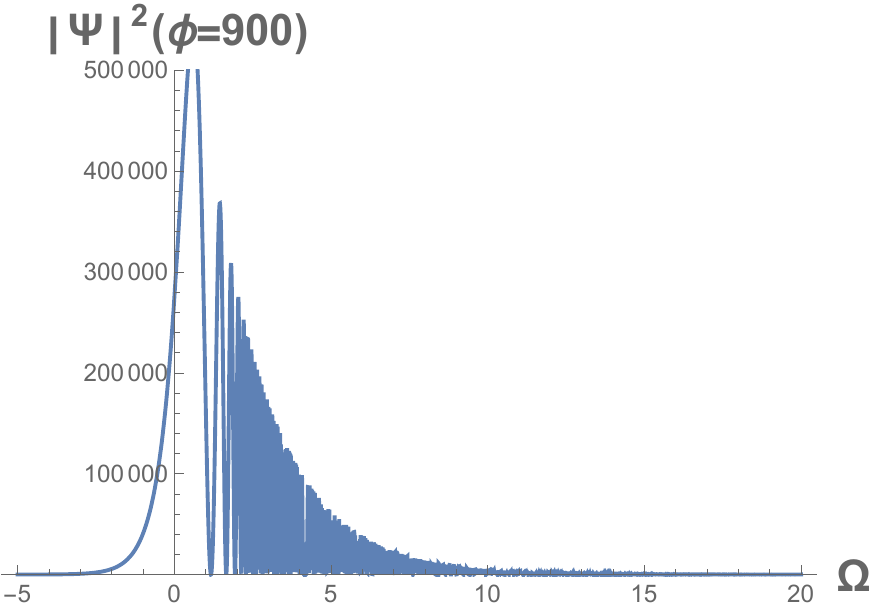}
\caption{These graphs
 represent a break-off for the defined value in
the scalar field $\phi$, for the case in the factor ordering Q = 1,
taking into account both solutions (\ref{densi}) given in Figure
(\ref{dust-era}). The first line corresponds to the solution with
the modified Bessel function, and the second line employs the
ordinary Bessel function. This behavior appears in a similar way for
different values in the factor ordering parameter, $ Q=-1,0,1$. We
can see that the values in the amplitude of the probability density
for big values in the scalar field, have very large growth,  acting
as a background in the classical level. } \label{profile-dust}
\end{figure}

Unlike the previous formalism, in~the present work, we employed a
barotropic equation with perfect fluid for the energy momentum
tensor in the K-essence scalar field into the Lagrangian and
Hamiltonian formalism, obtaining the momentum of the scalar field
with fractional numbers. However, the~momentum of the scale factor
appeared in the usual way. We obtained the classical solutions for
different scenarios in the Universe, which are similar to those
which were obtained for standard matter 16 years ago, see
Ref.~\cite{berbena} in Equations~(6) and (34) in the time $\tau$. In
this sense, we can introduce the idea that the kinetic energy of the
scalar field should falsify the standard matter by employing the
K-essence formalism. In~the quantum regime, we found a fractional
differential equation for the scalar field, where the
Mittag--Leffler function is the novel solution in many scenarios
with real or complex values in its argument z. With~this in mind, we
visualized two alternatives in our analysis; the first one is within
the traditional expectation over the behavior of the probability
density, where the best candidates for quantum solutions are
 those that have a damping behavior with respect to the scale factor, which appear in all scenarios under our study,
 without saying anything about
the scalar field. The~other alternative scenario is when we keep the
scale factor scenario, and~we consider the
 values of the scalar field as significant in the quantum regime, appearing in various scenarios in
the behavior of the Universe. This is mainly in those where the
Universe shows huge behavior, for~example, in~ the inflation-like
scenario, see Figure~\ref{inflation2} and the actual epoch,
 Figures~\ref{dust-era} or \ref{profile-dust},
 where the scalar field appears as a
 background. In~other words, the interpretation of probability density of the
unnormalized wave function, is given when we demand that $\Psi$ does
not diverge when the scale factor A (or $\Omega$) goes to infinity,
and the scalar field is arbitrary. However, the~ evolution with the
scalar field is now important in this class of theory and others,
as~it appears in some stages of evolution of our Universe, intended
to serve as a a background for the evolution of the Universe in the
classical world. The~quantum regime appears with big values in the
corresponding figures (see the corresponding  (\ref{dust-era}),
(\ref{profile-dust}) and (\ref{inflation2}) plots). However, it is
interesting to mention that in the radiation-like scenario,
Figures~\ref{radiation-era} and \ref{radiation-like}, this behavior
over the scalar field is less significant in the formation of atoms
and close to the stiff matter scenarios, where an oscillatory
behavior takes place in the scalar field. We briefly illustrate the
main results in this~work.

\begin{enumerate}
\item{} Using the K-essence formalism in a general way, applied to the Friedmann--Lema$\hat{\i}$tre--Robertson--Walker
cosmological model, we found the Hamiltonian density in the scalar
field momenta raised to a power with non-integers. This produces in
the quantum scheme a fractional differential equation  in a natural
way, such as in~this variable with order $\beta=\frac{2\alpha}{2
\alpha-1}$, where $\alpha \in (-1,\infty)$, which was solved for
different scenarios of our~Universe.

\item{} We found in the classical scheme that the time evolution $\tau$ of the scale factor
for ordinary matter was found  16 years ago by one of us; this time,
behavior is reproduced in the K-essence formalism, see
Equation~(\ref{scale-factor}) in this work, which is consistent with
the result obtained in the Ref.~\cite{berbena},  Equations~(6) and
(34), with~ordinary~matter.

\item{} In the quantum regime, the~novel solution at the fractional differential equation in the scalar field was found in
terms of the Mittag--Leffler function, with~a real or complex
argument, and we can see that this function appears in several
scenarios of our Universe in this work. This function is reported in
particular work dealing with different disciplines of~cosmology.

\item{} In one of our analyses presented on the probability density, we considered the
 values of the scalar field as significant in the quantum regime, appearing in various scenarios in
the behavior of the Universe; mainly in those where the Universe has
huge behavior. For~example, in~ the inflation-like scenario and the
actual epoch, where the scalar field appears as a background, the
quantum regime appears with big values, but~it presents a
 moderate development in other scenarios with a different ordering parameter Q.
\end{enumerate}

\vspace{6pt}
\appendix
\section{obtaining equations of motion. }\label{AppA}
We take the variation over the fields components in the action for
K-essence theory coupled with gravity
\begin{equation}
 S=\int \sqrt{-g}\left[ R + f(\phi){\cal G}(X)\right]d^4x,
\label{accion}
\end{equation}
where \emph{R} is the Ricci scalar, g is the determinant to the
metric, $ f(\phi)$ is a function of the scalar field, and~$ {\cal
G}[X]$ is a functional depending of the kinetic energy $
X(\phi,g^{\mu \nu})=-\frac{1}{2}g^{\mu \nu}\nabla_\mu \phi
\nabla_\nu \phi.$

 The variation of the fields $ (g^{\mu \nu},
\phi)$ in the action (\ref{accion}) becomes
\begin{eqnarray}
 \delta S &=& \int \delta \left[\sqrt{-g}R\right]d^4x+\int \delta
\sqrt{-g}\left[
f(\phi){\cal G}(X)\right]d^4x \nonumber\\
&&  +\int \sqrt{-g}\left[ \delta f(\phi){\cal G}(X)+ f(\phi)\delta
{\cal G}(X)\right]d^4x,\nonumber\\
&=& \int \sqrt{-g}\, G_{\lambda \theta} \delta g^{\lambda \theta}
d^4x + \int \sqrt{-g}\frac{1}{2}\left[- f(\phi){\cal
G}(X)\right]g_{\lambda \theta} \delta g^{\lambda \theta} d^4x
+\nonumber\\
&&  +\int \sqrt{-g}\left[ \frac{\partial f(\phi)}{\partial
\phi}\delta \phi{\cal G}(X)+ f(\phi)\frac{\partial {\cal
G}(X)}{\partial X}\delta X\right]d^4x,
\end{eqnarray}
where the variation of the functional  $ {\cal G}(X)$ is over the
kinetic energy
\begin{eqnarray}  \delta X(\phi)&=&
-\frac{1}{2}\nabla_\mu \phi \nabla_\nu \phi \delta g^{\mu \nu} -
\frac{1}{2}g^{\mu \nu}\nabla_\mu \delta \phi \nabla_\nu \phi
-\frac{1}{2}g^{\mu\nu}\nabla_\mu \phi
\nabla_\nu \delta \phi,\nonumber\\
&=& -\frac{1}{2}\nabla_\mu \phi \nabla_\nu \phi \delta g^{\mu \nu}
-\nabla^\nu \phi \nabla_\nu \delta \phi;\nonumber
\end{eqnarray}
introducing into the last equation, we have
\begin{eqnarray}
 \delta S &=&\int \sqrt{-g}\,\left[ G_{\lambda \theta} -
\frac{1}{2}f(\phi)\left({\cal G}(X)g_{\lambda \theta}+\frac{\partial
{\cal G}(X)}{\partial X}\nabla_\mu \phi \nabla_\nu \phi
\right)\right]
  \delta g^{\lambda \theta} d^4x \nonumber\\
 && +
 \int \sqrt{-g}\left[ \frac{\partial
f(\phi)}{\partial \phi}\delta \phi{\cal G}(X)+ f(\phi)\frac{\partial
{\cal G}(X)}{\partial X}\left\{  -\nabla^\nu \phi \nabla_\nu \delta
\phi\right\} \right]d^4x. \label{eq}
\end{eqnarray}

However, we know that the total derivative
\begin{eqnarray}
\nabla_\nu\left( f(\phi)G_X \nabla^\nu \phi \delta
\phi\right)&=&\frac{df(\phi)}{d\phi}\nabla_\nu \phi \nabla^\nu
\phi\, G_X \delta \phi+f(\phi) G_{XX} X_{,\nu} \nabla^\nu \delta
\phi +f(\phi) G_X\nabla_\nu^\nu \phi \delta \phi\nonumber\\
&& +f(\phi)G_X \nabla^\nu \phi \nabla_\nu \delta \phi.
\end{eqnarray}

Thus,
\begin{eqnarray}
-f(\phi)G_X \nabla^\nu \phi \nabla_\nu \delta \phi&=&
\frac{df(\phi)}{d\phi}\nabla_\nu \phi \nabla^\nu \phi\, G_X \delta
\phi+f(\phi) G_{XX} X_{,\nu} \nabla^\nu \delta \phi +f(\phi)
G_X\nabla_\nu^\nu \phi \delta \phi\nonumber\\
&& -\nabla_\nu\left( f(\phi)G_X \nabla^\nu \phi \delta \phi\right),
\end{eqnarray}
and reinserting into (\ref{eq}), we have
\begin{eqnarray}
 \delta S &=&\int \sqrt{-g}\,\left[ G_{\lambda \theta} -
\frac{1}{2}f(\phi)\left({\cal G}(X)g_{\lambda \theta}+\frac{\partial
{\cal G}(X)}{\partial X}\nabla_\mu \phi \nabla_\nu \phi
\right)\right]
 \delta g^{\lambda \theta} d^4x \nonumber\\
 && +
 \int \sqrt{-g}\left\{\frac{\partial
f(\phi)}{\partial \phi}\delta \phi{\cal
G}(X)+\frac{df(\phi)}{d\phi}\underbrace{\nabla_\nu \phi \nabla^\nu
\phi}_{-2X}\, G_X \delta \phi+f(\phi) G_{XX} X_{;\nu} \nabla^\nu
\phi \,
\delta \phi \right. \nonumber\\
&&\left. +f(\phi) G_X\nabla_{;\nu}^\nu \phi \delta \phi
-\nabla_\nu\left( f(\phi)G_X \nabla^\nu \phi \delta
\phi\right)\right\} d^4x, \nonumber\\
 &=&\int \sqrt{-g}\,\left[ G_{\lambda \theta} -
\frac{1}{2}f(\phi)\left({\cal G}(X)g_{\lambda \theta}+\frac{\partial
{\cal G}(X)}{\partial X}\nabla_\mu \phi \nabla_\nu \phi
\right)\right]
  \delta g^{\lambda \theta} d^4x \nonumber\\
 && +
 \int \sqrt{-g}\left\{\frac{df(\phi)}{d\phi}\left[ {\cal
G}(X)-2X G_X \right]+f(\phi) \left[G_{XX} X_{,\nu} \nabla^\nu +
G_X\nabla_\nu^\nu \phi\right] \right\} \delta \phi d^4x,
\end{eqnarray}
where we have eliminated the integral over the total~derivative.

  The
variation over the scalar field gives the equation of motion for
this field, being
\begin{equation}
\frac{df(\phi)}{d\phi}\left[ {\cal G}(X)-2X G_X \right]+f(\phi)
\left[G_{XX} X_{,\nu} \nabla^\nu + G_X\nabla_\nu^\nu \phi\right]=0,
\end{equation}
which corresponds to Equation~(\ref{Sfe}). For~obtaining the
Einstein field-like equations, we take the variation on the metric $
g^{\mu \nu}$,
\begin{equation}
G_{\mu \nu}=\frac{1}{2}f(\phi) \left[\nabla_\mu \phi \nabla_\nu \phi
\frac{\partial {\cal G}(X)}{\partial X}+ g_{\mu \nu}{\cal
G}(X)\right],
\end{equation}
where the energy-momentum tensor becomes
\begin{equation} T_{\mu \nu}(\phi)=
+\frac{1}{2}f(\phi) \left[\nabla_\mu \phi \nabla_\nu \phi
\frac{\partial {\cal G}(X)}{\partial X}+ g_{\mu \nu}{\cal
G}(X)\right],
\end{equation}
and considering the energy-momentum tensor of a barotropic perfect
fluid for the scalar~fields
\begin{equation}
 T_{\mu \nu}(\phi)=(\rho + P)u_\mu(\phi) u_\nu(\phi) + P\, g_{\mu
\nu},
\end{equation}
we have that the pressure P and the energy density $\rho$ of the
scalar fields become
\begin{equation}  P(\phi)=\frac{1}{2}f(\phi)
{\cal G} , \qquad \rho(\phi)=\frac{1}{2}f\left[ 2X \frac{\partial
{\cal G}}{\partial X}-{\cal G} \right],
\end{equation}
the four-velocity becomes $ u_\mu u_\nu=\frac{\nabla_\mu \phi
\nabla_\nu \phi}{2X}$ and the barotropic index $\omega_{X}$ is
\begin{equation}
 \omega_{X}=\frac{f(\phi) {\cal G}}{ f\left[ 2X \frac{\partial
{\cal G}}{\partial X}-{\cal G} \right]}. \label{barotro}
\end{equation}


\section{Obtaining the equations of motion with particular metric.}\label{AppB}
We have rewritten the line element in the time $\tau=Ndt=\prime$,
\begin{eqnarray}
ds^2&=&-N(t)^2 dt^2 + A^2(t) \left[dr^2 +r^2(d\theta^2+sin^2\theta
d\phi^2) \right], \label{frw-a}\\
&=&- d\tau^2 + A^2(\tau) \left[dr^2 +r^2(d\theta^2+sin^2\theta
d\phi^2) \right], \label{frw1-a}
\end{eqnarray}
where the metric element $g_{\tau \tau}=-1$ implies that
$\Gamma_{\tau \tau}^\tau=0$ and $\Gamma_{j\tau}^j=\Omega^\prime$,
$j=r,\theta,\phi=1,2,3$.

 When the function $f(\phi)$ is constant,
the Equation~(\ref{Sfe}) is reduced to
\begin{equation}
{\cal G}_X\phi^{,\nu}_{\, ;\nu} + {\cal G}_{XX}X_{;\nu}\phi^{,\nu}
=0; \label{21a}
\end{equation}
using the metric (\ref{frw1-a}), we obtain that the different
parameters into the Equation~(\ref{21a}) are
\begin{eqnarray}
X&=&\frac{1}{2}\left(\phi^\prime\right)^2, \quad \to \quad
\left(\phi^\prime\right)^2=2X, \qquad X^\prime=\phi^\prime
\phi^{\prime \prime}, \quad \to \quad
\phi^{\prime \prime}=\frac{X^\prime}{\phi^\prime}\nonumber\\
\phi^{,\nu}_{;\nu}&=&\phi^{,\nu}_{,\nu}+\Gamma_{\nu \rho}^\nu
\phi^\rho=\phi^{\prime \prime}+\left(\Gamma_{\tau \tau}^\tau
+\Gamma_{1 \tau }^1 +\Gamma_{2 \tau}^2 +\Gamma_{3 \tau }^3
\right)\phi^\prime=\phi^{\prime \prime}+3\Omega^\prime \phi^\prime.
\end{eqnarray}

Thus, the~Equation~(\ref{21a}) is rewritten as
\begin{equation}
{\cal G}_X \left(\phi^{\prime \prime}+3\Omega^\prime \phi^\prime
\right) + {\cal G}_{XX}X^\prime \phi^\prime ={\cal G}_X
\left(\frac{X^\prime}{\phi^\prime}+3\Omega^\prime \phi^\prime
\right) + {\cal G}_{XX}X^\prime \phi^\prime= 0; \label{21aa}
\end{equation}
multiplying by $\phi^{\prime}$, we have
\begin{equation}
 \left[{\cal
G}_X  + 2 X {\cal G}_{XX} \right]X^\prime + 6\Omega^\prime \, X
{\cal G}_X =0, \label{21final}
\end{equation}
where we had used the previous relations; this equation correspond
to (\ref{frw-k-model}).

Dividing between $XG_X$ the last equation, we have
\begin{eqnarray}
&& \frac{X^\prime}{X} + 2  \frac{{\cal G}_{XX}}{{\cal G}_X}\,
X^\prime + 6 \frac{A^\prime}{A} =\frac{d}{d\tau}\left(Ln X + Ln
{\cal G}_X^2+ Ln A^6\right) \nonumber\\
&&=\frac{d}{d\tau}Ln\left(A^6 X {\cal G}_X^2 \right)=0, \quad \to
\quad A^6 X {\cal G}_X^2=\eta=constant,
\end{eqnarray}
obtaining that
\begin{equation}
X {\cal G}_X^2=\eta A^{-6}, \label{b8}
\end{equation}
which is the Equation~(\ref{frw-k-solution}).

When ${\cal G}=X^\alpha$ and substituting into (\ref{b8}), we have
$\alpha^2 X^{2\alpha-1}=\eta A^{-6}$ with
$X=\frac{1}{2}\left(\frac{d\phi}{d\tau}\right)^2$, obtaining for the
scalar field $\phi$ the equation
\begin{equation} \frac{d\phi}{d\tau}=
\sqrt{2}
\left[\left(\frac{\eta}{\alpha^2}\right)^{\frac{1}{2(2\alpha-1)}}
A^{-\frac{3}{2\alpha-1}}\right]=\sqrt{2}
\left[\left(\frac{\eta}{\alpha^2}\right)^{\frac{1}{2(2\alpha-1)}}
e^{-\frac{3}{2\alpha-1}\Omega}\right],\label{a-phi-tau}
\end{equation}
 whose solution in the time $\tau$ is
\begin{equation}
 \Delta \phi=\sqrt{2}
{\left(\frac{\eta}{\alpha^2}\right)^{\frac{1}{2(2\alpha-1)}}} \int
A^{-\frac{3}{2\alpha-1}} d\tau =\sqrt{2}
{\left(\frac{\eta}{\alpha^2}\right)^{\frac{1}{2(2\alpha-1)}}}  \int
e^{-\frac{3}{2\alpha-1}\Omega} d\tau, \label{a-ss1}
\end{equation}

\section{Equivalence between Lagrangian densities}\label{AppC}

The canonical Lagrangian density ${\cal L}_{canonical}(q_i,
\Pi_i,t)$ (\ref{canonical}) in gravitation theories is obtained from
the usual Lagrangian density ${\cal L}(q_i,\dot q_i,t)$
(\ref{lala}), rewritten the velocities $\dot q_i$ in term of the
momenta $\Pi_i=\frac{\partial {\cal L}}{\partial \dot q_i}$ to the
corresponding coordinate field $q_i$. With~this procedure,
the~canonical Lagrangian density appear directly written as a
Lagrangian density in constrained systems, where the Lagrangian
multiplier is the lapse function $N(t)$, being the corresponding
gauge parameter in this theory. This is equivalent to using the
canonical transformation where the hamiltonian density is ${\cal
H}=\Pi_j \dot q^j - {\cal L}$. However, from~this point of view,
in~this expression the ${\cal H}$ must be interpreted as ${\cal H}=N
{\cal H}_{canonical}$ where the lapse function $N$ appears as a
lagrangian multiplier. In~the following we realize this calculation,
employing the usual canonical transformation.
 We have the momenta
\begin{eqnarray}
\Pi_\Omega&=& \frac{12}{N}e^{3\Omega}\dot \Omega, \quad \rightarrow \quad \dot \Omega=\frac{N}{12}e^{-3\Omega}\Pi_\Omega, \label{abc} \\
 \Pi_\phi&=&
-\left(\frac{1}{2}\right)^\alpha\frac{2\alpha}{N^{2\alpha-1}}e^{3\Omega}{\dot
\phi}^{2\alpha -1}, \quad \rightarrow \quad \dot
\phi=-N\left[\frac{2^\alpha}{2\alpha}e^{-3\Omega}\Pi_\phi\right]^{\frac{1}{2\alpha
-1}}, \label{pi-a}
\end{eqnarray}
substituting into the canonical transformation between the
Hamiltonian density and Lagrangian density $${\cal H}=\Pi_j \dot q^j
- {\cal L}$$
\begin{eqnarray}
{\cal H}&=&\Pi_\Omega \dot \Omega + \Pi_\phi \dot \phi - e^{3\Omega}
\left[6 \frac{\dot \Omega^2 }{N}
  - \left(\frac{1}{2}\right)^{\alpha}\left( \dot \phi\right)^{2\alpha} N^{-2\alpha+1}
  \right]\nonumber\\
&=&\Pi_\Omega\left(
\frac{N}{12}e^{-3\Omega}\Pi_\Omega\right)+\Pi_\phi
\left(-N\left[\frac{2^\alpha}{2\alpha}e^{-3\Omega}\Pi_\phi\right]^{\frac{1}{2\alpha
-1}} \right)\nonumber\\
&&\quad -e^{3\Omega} \left[\frac{6}{N} \left(
\frac{N}{12}e^{-3\Omega}\Pi_\Omega,\right)^2
  - \left(\frac{1}{2}\right)^{\alpha}\left(  -N\left[\frac{2^\alpha}{2\alpha}e^{-3\Omega}\Pi_\phi\right]^{\frac{1}{2\alpha
-1}}  \right)^{2\alpha} N^{-2\alpha+1}
  \right]\nonumber\\
  &=&Ne^{3\Omega}\Pi_\Omega^2
  \left(\frac{1}{12}-\frac{1}{24}\right)
  -Ne^{-\frac{3\Omega}{2\alpha-1}}\left(\frac{2^{\alpha-1}}{\alpha}\right)^{\frac{1}{2\alpha-1}}\Pi_\phi^{\frac{2\alpha}{2\alpha-1}}\nonumber\\
&&  +Ne^{-\frac{3\Omega}{2\alpha-1}}
\frac{1}{2\alpha}\left(\frac{2^{\alpha-1}}{\alpha}\right)^{\frac{1}{2\alpha-1}}
  \Pi_\phi^{\frac{2\alpha}{2\alpha-1}}\nonumber\\
  &=&N\frac{e^{-3\Omega}}{24}\Pi_\Omega^2 -Ne^{-\frac{3\Omega}{2\alpha-1}}\left(\frac{2^{\alpha-1}}{\alpha}\right)^{\frac{1}{2\alpha-1}}
  \left[\frac{1}{1-2\alpha} \right]\Pi_\phi^{\frac{2\alpha}{2\alpha-1}}\nonumber\\
  &=&N\frac{e^{-3\Omega}}{24}\Pi_\Omega^2 -Ne^{-\frac{3\Omega}{2\alpha-1}}\,\frac{2\alpha-1}{2\alpha}
  \left(\frac{2^{\alpha-1}}{\alpha}\right)^{\frac{1}{2\alpha-1}}\Pi_\phi^{\frac{2\alpha}{2\alpha-1}}\nonumber\\
  &=& N\frac{e^{-\frac{3\Omega}{2\alpha-1}}}{24}\left(e^{-\frac{6(\alpha-1)}{2\alpha-1}\Omega}\Pi_\Omega^2
  -\frac{12(2\alpha-1)}{\alpha}  \left(\frac{2^{\alpha-1}}{\alpha}\right)^{\frac{1}{2\alpha-1}} \Pi_\phi^{\frac{2\alpha}{2\alpha-1}}
  \right), \label{equiv}
 \end{eqnarray}
  corresponding to Equation~(\ref{hami}).

\bigskip
 \noindent Author Contributions:  Conceptualization, J.S,
J.J.R; Methodology, J.S,  J.J.R;  Writing-Original Draft, J.S,
J.J.R; Writing-Review and Editing, J.S,  J.J.R; Visualization, J.S,
J.J.R.
\bigskip

\noindent All authors have read and agreed on the published version
of the manuscript.
\bigskip

\noindent Funding: J.S. was partially supported by PROMEP grants
UGTO-CA-3. Both authors were partially supported by SNI-CONACyT. J.J.
Rosales is supported by the UGTO-CA-20 nonlinear photonics and
Department of Electrical Engineering.
\bigskip

\noindent Data Availability Statement: Not applicable.
\bigskip

\noindent Acknowledgments: Authors thank anonymous referees, since
by answering their criticisms we understood our problem better. This
work is part of the collaboration within the Instituto Avanzado de
Cosmolog\'ia and Red PROMEP: Gravitation and Mathematical Physics,
under project Quantum aspects of gravity in cosmological models,
phenomenology, and geometry of space-time. Many calculations were
done by Symbolic Program REDUCE 3.8.
\bigskip

\noindent Conflicts of Interest: The authors declare no conflict of
interest.

\end{document}